\documentclass[sigconf]{acmart}

\AtBeginDocument{%
  \providecommand\BibTeX{{%
    \normalfont B\kern-0.5em{\scshape i\kern-0.25em b}\kern-0.8em\TeX}}}

\newcommand{\CHANGE}[1]{\textcolor{black}{#1}}

\usepackage{microtype}

\newcommand{\PROMPT}[1]{%
    \begingroup
    \spaceskip=0.3em plus 0.1em minus 0.1em 
    \textsf{\textcolor{black}{\small{\fontfamily{cmss}\selectfont ``#1''}}}%
    \endgroup
}

\copyrightyear{2024}
\acmYear{2024}
\setcopyright{acmlicensed}\acmConference[UIST '24]{The 37th Annual ACM Symposium on User Interface Software and Technology}{October 13--16, 2024}{Pittsburgh, PA, USA}
\acmBooktitle{The 37th Annual ACM Symposium on User Interface Software and Technology (UIST '24), October 13--16, 2024, Pittsburgh, PA, USA}
\acmDOI{10.1145/3654777.3676375}
\acmISBN{979-8-4007-0628-8/24/10}

\usepackage{hhline}
\usepackage{multirow}
\usepackage{makecell}
\usepackage{bm}
\usepackage{algorithm}
\usepackage{algorithmic}
\usepackage[T1]{fontenc}
\usepackage{xcolor,colortbl}
\usepackage{makecell}
\usepackage{array}
\usepackage{amsmath}
\usepackage{makecell}
\usepackage{float}

\usepackage{graphicx}
\usepackage{dblfloatfix} 
\usepackage{placeins}

\newcolumntype{P}[1]{>{\centering\arraybackslash}p{#1}}
\newcolumntype{M}[1]{>{\centering\arraybackslash}m{#1}}
\newcolumntype{L}[1]{>{\arraybackslash}p{#1}}
\newcolumntype{Z}[1]{>{\raggedright\let\newline\\\arraybackslash\hspace{0pt}}m{#1}}

\def\name{WorldScribe}

\author{Ruei-Che Chang}
\email{rueiche@umich.edu}
\affiliation{
 \institution{University of Michigan}
 \city{Ann Arbor, MI}
 \country{USA}
}
\author{Yuxuan Liu}
\email{liurick@umich.edu}
\affiliation{
 \institution{University of Michigan}
 \city{Ann Arbor, MI}
 \country{USA}
}
\author{Anhong Guo}
\email{anhong@umich.edu}
\affiliation{
 \institution{University of Michigan}
 \city{Ann Arbor, MI}
 \country{USA}
}
\begin{document}

\title{{\name}: Towards Context-Aware Live Visual Descriptions
}

\renewcommand{\shortauthors}{Ruei-Che Chang, Yuxuan Liu, and Anhong Guo.}

\begin{abstract}
Automated live visual descriptions can aid blind people in understanding their surroundings with autonomy and independence. 
However, providing descriptions that are rich, contextual, and just-in-time has been a long-standing challenge in accessibility. 
In this work, we develop \textit{WorldScribe}, a system that generates automated live real-world visual descriptions that are customizable and adaptive to users' contexts:
\textit{(i)} WorldScribe's descriptions are tailored to users' intents and prioritized based on semantic relevance.
\textit{(ii)} WorldScribe is adaptive to visual contexts, \textit{e.g.,} providing consecutively succinct descriptions for dynamic scenes, while presenting longer and detailed ones for stable settings. 
\textit{(iii)} WorldScribe is adaptive to sound contexts, \textit{e.g.,} increasing volume in noisy environments, or pausing when conversations start.
Powered by a suite of vision, language, and sound recognition models, WorldScribe introduces a description generation pipeline that balances the tradeoffs between their richness and latency to support real-time use.
The design of WorldScribe is informed by prior work on providing visual descriptions and a formative study with blind participants. 
Our user study and subsequent pipeline evaluation show that WorldScribe can provide real-time and fairly accurate visual descriptions to facilitate environment understanding that is adaptive and customized to users' contexts. 
Finally, we discuss the implications and further steps toward making live visual descriptions more context-aware and humanized.
\end{abstract}



\begin{teaserfigure}
 \vspace{-1pc}
 \includegraphics[width=\linewidth] {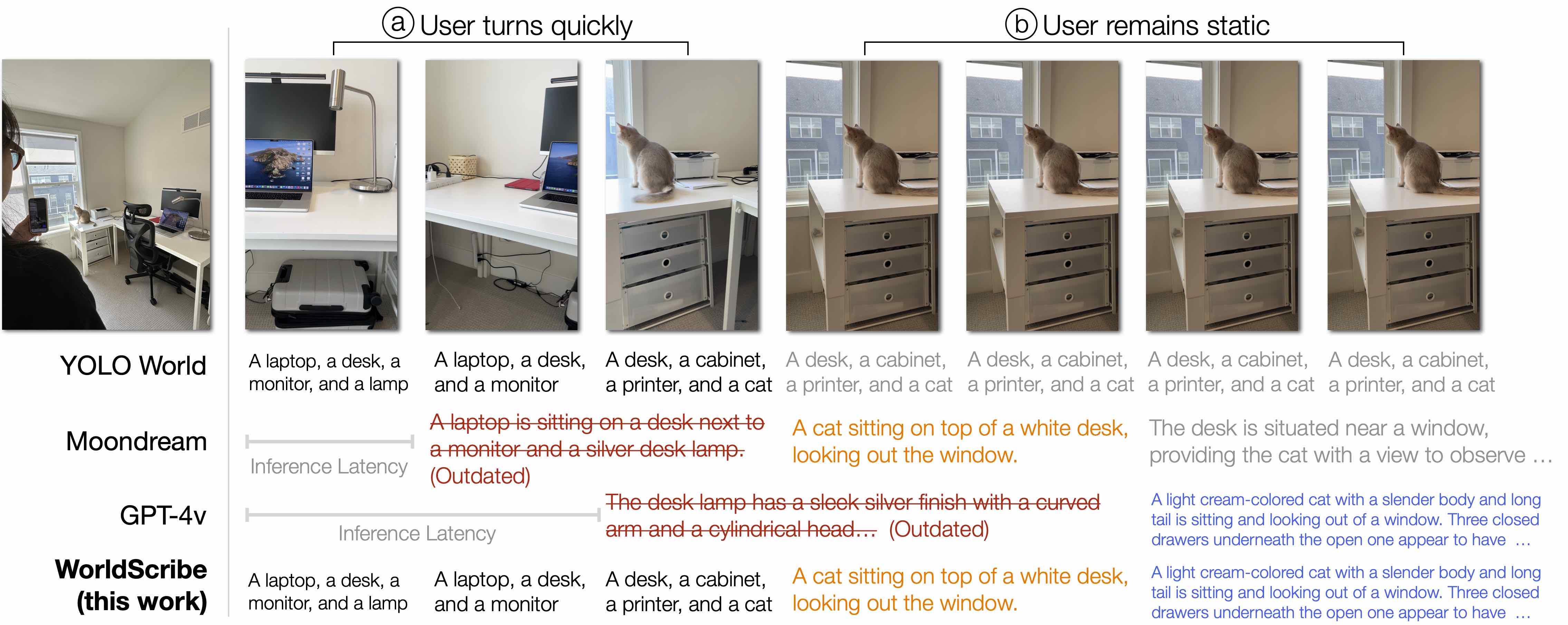}
 \vspace{-1.8pc}
 \caption{{\name} towards making the real world accessible for blind people through context-aware live visual descriptions. {\name} dynamically combines different vision-language models to provide live adaptive descriptions.
 (a) When the user turns quickly to scan the environment and yields frequent visual changes, {\name} generates basic descriptions with word-level labels (\textit{e.g.,} YOLO World \cite{yolow}) or general descriptions with objects and spatial relationships (\textit{e.g.,} Moondream \cite{moondream}). 
 On the other hand,
 (b) when the user remains static and faces a new scene for a duration that indicates their interests, {\name} provides rich descriptions from an overview to details (\textit{e.g.,} GPT-4v \cite{gpt4v}) to facilitate their visual scene understanding.
 }
 \Description{Figure 1
There are seven images in the figure in total. The first three describe figure 1 a that user turns quickly and the remaining four images describe figure 1 b that user remains static. Under the images there are three four rows represent descriptions generated by four different models, from top to bottom including YOLO World, Moondream, GPT-4v and WorldScribe (this work). 
In the row of YOLO World, there are descriptions from left to right images: (1) A laptop, a desk, a monitor, and a lamp. (2)  A laptop, a desk, and a monitor. (3) A desk, a cabinet, a printer, and a cat. (4) A desk, a cabinet, a printer, and a cat. This sentence is grayed out, representing not used in WorldScribe. (5) A desk, a cabinet, a printer, and a cat. This sentence is grayed out, representing not used in WorldScribe. (6) A desk, a cabinet, a printer, and a cat. This sentence is grayed out, representing not used in WorldScribe. (7) A desk, a cabinet, a printer, and a cat. This sentence is grayed out, representing not used in WorldScribe. 
In the row of Moondream, there are descriptions from left to right: (1) no description. And there is a bar to indicate there is a latency of model inference, stretching for one image length. (2) A laptop is sitting on a desk next to a monitor and a silver desk lamp (outdated). This sentence has strikethrough and is red, representing this description is outdated based on the current scene. This sentence stretch across two images as this sentence is longer. (2) A cat sitting on top of a white desk, looking out the window. This sentence is orange and is picked up by the WorldScribe used as follows. This sentence stretch across two images as this sentence is also longer.  (3) The desk is situated near a window, providing the cat with a view ot observe…. This sentence stretch across two images as this sentence is longer. This sentence is grayed out, representing not used in WorldScribe. 
In the row of GPT-4v, there are descriptions from left to right: (1) no description. And there is a bar to indicate there is a latency of model inference, stretching for two images length which is more than Moondream as GPT-4v takes longer inference time. (2) The desk lamp has a sleek silver finish with a curved arm and a cylindrical head… (Outdated) This sentence has strikethrough and is red, representing this description is outdated based on the current scene. This sentence also stretch to three images, indicating its longer sentence. (3) A light cream-colored cat with a slender body and long tail is sitting and looking out of a window. Three closed drawers underneath the open one appear to have  …. This sentence stretches for two images.
In the row of WorldScribe (this work), there are descriptions from left to right: (1) A laptop, a desk, a monitor, and a lamp. This sentence stretches for one image, borrowed from Yolo World. (2) A laptop, a desk, and a monitor. This sentence stretches for one image, borrowed from Yolo World. (3) A desk, a cabinet, a printer, and a cat. This sentence stretches for one image, borrowed from Yolo World. (4)  A cat sitting on top of a white desk, looking out the window. This sentence is orange and borrowed from Moondream, and stretches for two images as well. (5) A light cream-colored cat with a slender body and long tail is sitting and looking out of a window. Three closed drawers underneath the open one appear to have  ….  This sentence stretches for two images, borrowed from GPT-4v.
}
 \vspace{0pc}
 \label{fig:teaser}
\end{teaserfigure}

\maketitle

\section{Introduction}
Automated live visual descriptions can help blind or visually impaired (BVI) people understand their surroundings autonomously and independently. 
Imagine Sarah, who is blind, is exploring the zoo with her 2-year-old toddler. 
As they walk around the African grassland section, live visual descriptions provide rich information about the texture of the terrain animals are resting on and occasionally notify her about the movement of the zebras and rhinos. 
They join a giraffe feeding tour, and live visual descriptions narrate when a couple of giraffes reach out near her toddler. She feeds the lettuce leaves in her hand to them and snaps a nice photo.
Such contextual visual descriptions can supplement their environmental understanding and support a range of ever-changing scenarios. 

However, providing rich and contextual descriptions has been a long-standing challenge in accessibility. Researchers have explored ways to provide BVI individuals with visual descriptions across various visual media. 
For example, traditional AI captioning for digital media (\textit{e.g.}, images, videos) offers basic information but often misses the nuanced details BVI users need in varied contexts \cite{stangl21assets,kreiss2022context}.
While human-powered \cite{vizwiz, youdescribe, viscene, noviceauthoring} or human-AI hybrid techniques \cite{twittera11y, rescribe, omniscribe} deliver more detailed descriptions asynchronously for digital media, they fall short in real-world scenarios that require descriptions to be timely and pertinent to the user's context.
As a result, existing solutions in describing the real world have been limited to leveraging remote sighted assistance (RSA) to access BVI users' live camera streams and describe the visuals, such as Chorus:View \cite{chorusview} with crowd workers, BeMyEyes \cite{bemyeyes} with volunteers, and Aira \cite{aira} with trained agents. 
However, these human services can be extremely costly, not always available, and potentially raise privacy concerns. 

The advent of vision-language models (VLMs) and large language models (LLMs) makes it possible to provide automated visual descriptions without human assistance. 
Off-the-shelf tools, such as SeeingAI \cite{seeingai}, EnvisionAI \cite{envisionai} or ImageExplorer \cite{imageexplorer}, enable BVI people to upload an image and receive detailed descriptions.
However, \CHANGE{the asynchronous and one-size-fits-all nature of the produced descriptions makes it difficult to adapt these tools to dynamic real-world scenarios.}
Providing seamless real-time automated descriptions is further challenging when considering user needs and contexts \cite{stangl21assets,kreiss2022context}.
For instance, BVI people have individual preferences \cite{stangl21assets} (\textit{e.g.,} different visual experiences and familiarity with the environment), and the rich visual contexts may influence information priority depending on users' needs (\textit{e.g.,} walking on the street, or visiting a museum).
Furthermore, real-world sounds could hinder the perception of spoken descriptions \cite{soundshift}.
Therefore, when providing live visual descriptions in the real world, it is crucial to collectively consider \textit{user intent, visual and sound contexts} \CHANGE{(which we will refer to as the user's context throughout the paper)}.

\begin{figure}[t]
\begin{center}
\vspace{-1pc}
\includegraphics[width=\linewidth]{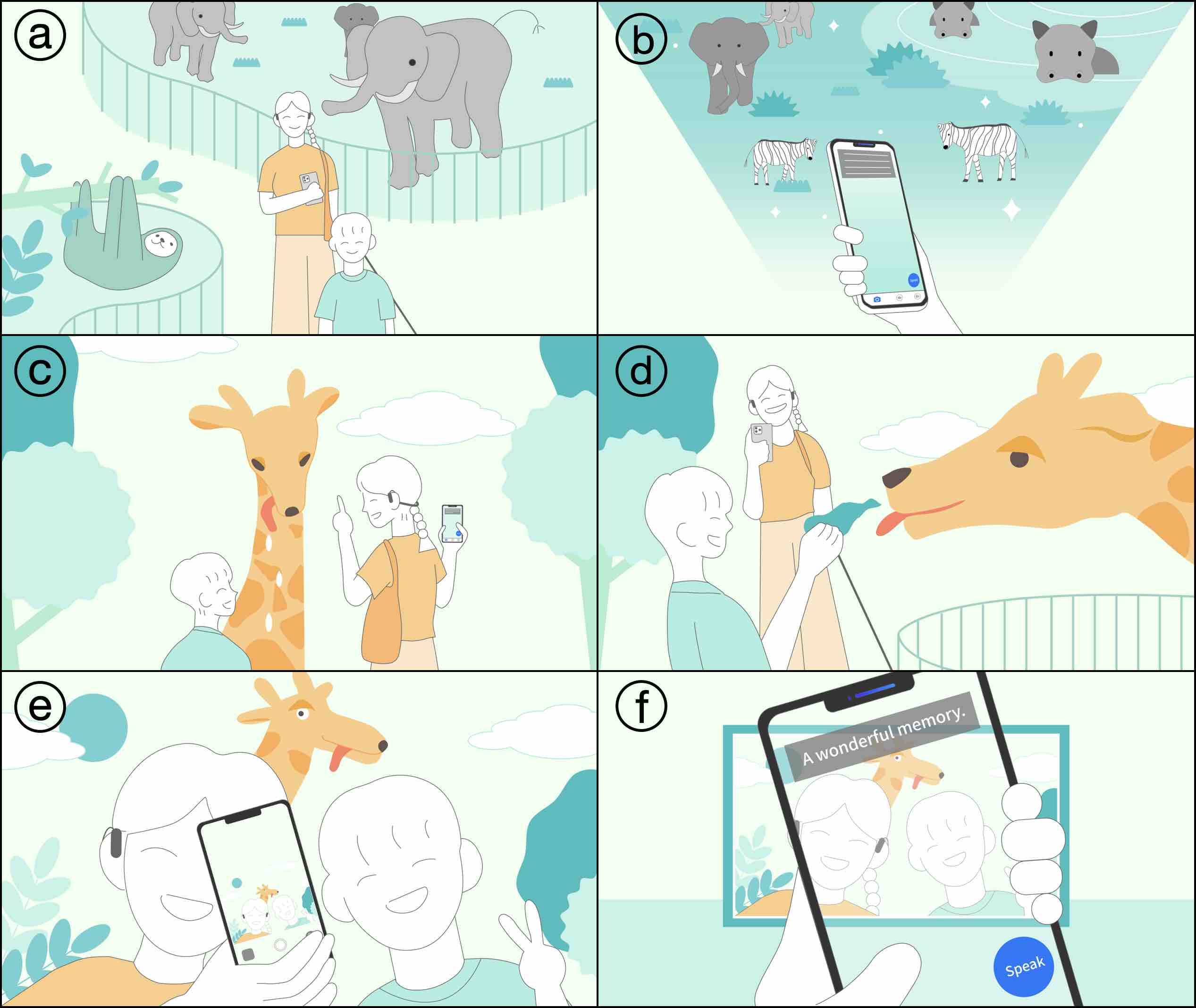}
\vspace{-1.5pc}
\caption{
(a) Sarah is exploring the zoo with her toddler using {\name}, (b) which  
describes surroundings to her. 
(c, d) They join a giraffe feeding tour, live visual descriptions narrate when giraffes reach out near her toddler, who feeds the lettuce leaves to them and (e, f) snaps a nice photo.
}
\vspace{-1.5pc}
\label{fig:zoo}
\Description{Figure 2
Figure 2 a 
Sarah is exploring the zoo with her toddler using WorldScribe. The image shows one woman holding a cane is walking with a child in the zoo. The zoo has several animals such as elephants and sloth.
Figure 2 b 
WorldScribe describes her surroundings to her. The image shows a projection from the phone, which contains several animals like elephants, rhinos, zebras, and grassland. 
Figure 2 c,d 
They join a giraffe feeding tour, live visual descriptions narrate when giraffes reach out near her toddler, who feeds the lettuce leaves to them. The two images show that the woman and the child are interacting with a giraffe and the child is feeding a giraffe.
Figure 2 e,f
They snap a nice photo. The images show the woman holding the phone and taking selfies with the child. 
}
\end{center}
\end{figure}

Informed by prior works on providing visual descriptions and a formative study with five BVI people, we identified design considerations for a system to provide live visual descriptions in the real world,
such as providing descriptions with overview first then adaptive details on the fly, prioritizing descriptions based on semantic relevance, and enabling customizability based on varied user needs. 
We then develop \textit{{\name}}, a system that generates automated visual descriptions that are adaptive to the users' contexts in real-time, and in the real world. 
First, {\name} is adaptive to the user's intent, \textit{e.g.,} prioritizing the most pertinent descriptions based on semantic relevance, and visual attributes based on user customizations.
Second, {\name} is adaptive to visual contexts; for instance, it provides consecutively succinct descriptions for dynamic visual scenes while it presents longer and more detailed ones for stable settings.
Third, {\name} is also adaptive to sound contexts, \textit{e.g.,} increasing description volume in noisy environments, or pausing when conversations start. 

Powered by a suite of vision, language, and sound recognition models,
{\name} introduces a description generation pipeline with different VLMs that balances the tradeoffs between their richness and latency to support real-time usage (Figure \ref{fig:teaser}). 
{\name} dynamically assigns prompts to VLMs encoded with user customizations on their information needs and in-the-moment visual contexts (\textit{e.g.,} busy or static), and prioritizes descriptions based on the user intent and the proximity of the described content to the user. 
{\name} also keeps the spoken descriptions up-to-date by examining the object compositions and similarity between the VLM-referred and current camera frames and the changes in user orientations.

Our pipeline evaluation shows that {\name} can provide fairly accurate visual descriptions, cover important information, and prioritize descriptions based on users' intent and proximity. 
Furthermore, our user study with six BVI participants demonstrates that {\name} enables effective environment understanding that is adaptive and customizable to users' contexts. 
However, there is still a gap in making AI-generated descriptions humanized, user-centric, and context-aware, which we discuss and provide implications for future work. 
{\name} represents an important step towards solving this long-standing accessibility challenge, and its technical approach may find applications broadly for enhancing real-time visual assistance to promote real-world and digital media accessibility.

\begin{table*}[t]
\caption{Overview of research or commercial apps for describing visual media. AD denotes audio description.}
\label{tab:apps}
\vspace{-1pc}
\renewcommand{\arraystretch}{1.2}
\begin{tabular}{Z{2.2cm} Z{2.5cm} Z{2.4cm} M{1.4cm} M{0.5cm} Z{2.9cm} Z{3.6cm}}

\Xhline{3\arrayrulewidth}
\rowcolor{gray!20} 
\textbf{App Category} & \textbf{Application} & \textbf{Description type}   & \textbf{Enabling source} & \textbf{Real time} & \textbf{Audio presentation} & \textbf{Customization options}  \\ 
\Xhline{3\arrayrulewidth}

\multirow{3}{=}{Navigation} 
& BlindSquare \cite{blindsquare} & Audio direction & Map & \checkmark & Spatial audio & Landmarks \\ 
\cline{2-7}
& SoundScape \cite{soundscape} & Audio direction & Map & \checkmark & Spatial audio & Landmarks, route \\ 
\cline{2-7}
& NavCog \cite{navcog} & Audio direction & Map & \checkmark & & Landmarks \\ 
\Xhline{2\arrayrulewidth}

\multirow{3}{=}{Image Understanding} 
& Seeing AI \cite{seeingai} & Image description & AI & & & Short or detailed content \\
\cline{2-7}
& Envision AI \cite{envisionai} & Image description & AI & & & Short or detailed content \\ \cline{2-7}
& ImageExplorer \cite{imageexplorer} & Image description & AI & & & Number of info layers, accuracy, specific object info \\ 
\Xhline{2\arrayrulewidth}

\multirow{3}{=}{Video Understanding} 
& YouDescribe \cite{youdescribe} & Audio description & Human & & inline ADs &  \\ 
\cline{2-7}
& Rescribe \cite{rescribe} & Audio description & Human-AI & & inline ADs & \\ 
\cline{2-7}
& OmniScribe \cite{omniscribe} & Audio description & Human-AI & & Spatial Audio, inline and extended ADs & \\ \cline{2-7}
& ShortScribe \cite{shortscribe} & Audio description & AI & & & \\ 
\cline{2-7}
& InfoBot \cite{narratebot} & Video VQA & AI & & & Request to AI \\ 
\Xhline{2\arrayrulewidth}

\multirow{2}{=}{Remote Sighted Assistance} 
& Aira \cite{aira} & Verbal Guidance & Human & \checkmark & & Request to sighted agents \\ 
\cline{2-7}
& BeMyEyes \cite{bemyeyes} & Verbal Guidance & Human & \checkmark & & Request to sighted agents \\ 
\Xhline{2\arrayrulewidth}

\multirow{2}{=}{Visual Question Answering} 
& VizWiz \cite{vizwiz} & Image description & Human & & & Request to crowd workers \\ 
\cline{2-7}
& BeMyAI \cite{bemyai} & Image description & AI &  & & Request to AI \\

\Xhline{3\arrayrulewidth}
\textbf{Real-world Visual \newline Understanding} & \textbf{\name} \newline \textbf{(this work)} & \textbf{Live visual description} & \textbf{AI} & \textbf{\checkmark} & \textbf{Auto volume adjustment or pause} & \textbf{User intents, objects, visual attributes, audio presentations, verbosity} \\ 
\Xhline{3\arrayrulewidth}

\end{tabular}
\end{table*}

\section{Related work}
Our work builds upon prior work to provide BVI people with descriptions for accessing digital media and the real world, in order to fulfill their diverse needs. We describe our motivation and insights from previous literature below.

\subsection{Descriptions for Digital Visual Media}\label{RW_digital}
To understand digital visual media, BVI people typically rely on textual descriptions. 
World Wide Web Consortium has established Web Content Accessibility Guidelines for creators to add proper captions to images \cite{w3cimageconcepts} and audio descriptions to videos \cite{adguideline, W3CextendedAD, WCAG_AD} for BVI people to receive equal information as sighted people.
\CHANGE{
Several platforms allowed BVI people to request descriptions for images and videos that lack accessible visual descriptions from volunteer describers, such as YouDescribe \cite{youdescribe} and VizWiz \cite{vizwiz}. 
}
Despite the availability of these resources, learning those guidelines and providing good descriptions remain difficult \cite{viscene}; the scarcity of human resources also makes it hard to address the high volume of requests from BVI people \cite{hidead}, who may have different information needs based on their access contexts \cite{stanglchi20,stangl21assets}. 

\CHANGE{
To address these challenges, semi-automatic AI systems have been developed to streamline the description authoring process, such as generating initial image captions \cite{twittera11y} or audio descriptions \cite{narratebot, narrationbot2020, rescribe}. 
Although these systems reduce laborious tasks, they still require human effort to make one-size-fits-all descriptions usable.
Recently, VLM-powered systems can generate high-quality audio descriptions comparable to human describers \cite{shortscribe} and allow BVI people to query visual details interactively \cite{narratebot, genassist, editscribe, bemyai}.
However, the asynchronous and one-size-fits-all nature of descriptions makes it difficult to adapt these tools to dynamic real-world scenarios.
In response, this work aims to provide live contextual descriptions for BVI users by understanding their intent and visual contexts.
This is achieved through a description generation pipeline with dynamic prompt assignments based on user contexts, and different VLMs that balance the tradeoffs between their richness and latency to achieve real-time purposes.}

\subsection{Descriptions for Real-World Accessibility} \label{RW_realworld}
Accessing the real world through descriptions enhances BVI individuals' independence in various tasks, such as object identification \cite{recog, objecttrain, findmything}, line following \cite{linechaser}, and navigation \cite{bbeep, corridor, blindpilot, pathfinder, engel, navcog, blindsquare, soundscape}.
Navigation is especially important but challenging in unfamiliar settings, which demands extensive environmental understanding \cite{pathfinder,engel}, such as recognizing intersections \cite{corridor, pathfinder}, signs \cite{pathfinder, afif2020recognizing, yamanaka2021one}, and traffic light statuses \cite{chen2020traffic, tan2021flying}.
While these systems offered critical task-specific guidance (\textit{e.g.,} audio directions), they lacked visual descriptions for ever-changing surroundings.
Tools, such as SeeingAI \cite{seeingai}, ImageExplorer \cite{imageexplorer}, and Envision AI \cite{bemyai}, enabled BVI users to snap a photo and receive comprehensive visual descriptions within seconds, while BeMyAI \cite{bemyai} allowed BVI people to access details through turn-by-turn interactions (Table \ref{tab:apps}).
Yet, their utility falls short in rapidly changing visual scenes that require live and continuous descriptions.

An alternative for understanding the dynamic real world is through human assistance, such as RSA, which connects BVI users with sighted agents via video calls to fulfill requests through verbal guidance.
However, conveying visual information in this way can be challenging and cognitively demanding \cite{rsachallengechi20, kamikubo},
where agents were under pressure to understand and effectively communicate key details \cite{rsasurvey,lee2018conversations,rsachallengechi20}, or they needed to tailor the level of detail to the user's needs \cite{holmes2015iphone,rsachallengechi20}.
Moreover, RSA services could raise privacy concerns \cite{socialvizwiz}, incur high costs (\textit{e.g.,} \$65 for 20 monthly minutes with professional services, such as Aira \cite{aira}), and volunteer-based options, such as BeMyEyes \cite{bemyeyes}, may not always be available.
In this work, we aim to make the real world accessible to BVI people through automated live visual descriptions in order to enhance their environmental understanding beyond navigation instructions.

\subsection{Fulfilling Diverse Needs of BVI People}\label{RW_varied}
Creating high-quality descriptions that meet BVI people's diverse information needs is challenging for different visual media.
Prior research found that current one-size-fits-all approaches to image descriptions are insufficient for providing necessary details for meaningful interaction \cite{stanglchi20,stangl21assets,morris2018chi,salisbury2017toward,salisbury2018evaluating,kreiss2022context}.
Stangl et al. \cite{stanglchi20,stangl21assets} identified that the source of an image and the user's information goal impacted their information wants, and proposed universal terms (\textit{e.g.,} having identity or names for describing people) as minimum viable content, with other terms provided on demand based on users' contexts (\textit{e.g.,} person's height, hair color, etc.).
Guidelines also indicated that the inclusion of certain visual details should be context-based, such as having general information for first access or having details (\textit{e.g.,} color, orientations of objects) when gauging one's understanding of certain image content \cite{specificadguidelines,petrie2005describing}.

The varied information needs of BVI people were also found when accessing different types of videos \cite{narratebot,customad,jiang2024s}, such as different preferences on the audio description content (\textit{e.g.,} object details, settings), and output modalities (\textit{e.g.,} audio, tactile).
For 360-degree videos, which offer richer visual information and immersion, BVI people also have varied preferences for linguistic aspects (\textit{e.g.,} level of details, describing from first- or second-person view), or audio presentations (\textit{e.g.,} spatial audio) \cite{jiang2023assets,omniscribe}.
Thus, the way of presenting visual information and determining its richness is crucial and depends on the user's needs and context.

These findings of users' varied preferences for digital media also extended to the real world. 
Herskovitz et al. \cite{diyat} identified the needs of BVI individuals, who often customize assistive technologies for daily activities by combining mobile apps for different visual tasks, such as obtaining clock directions from Compass or descriptions from BlindSquare, or filtering visual information (\textit{e.g.,} text or colors).
Overall, prior work in both the digital and real worlds has highlighted the importance of customization for adapting to diverse user preferences and contexts. 
In this work, we explore live visual descriptions that are context-aware, by enabling customization options on the description content (\textit{e.g.,} level of details, visual attributes), and audio presentation (\textit{e.g.,} pausing or increasing volume) to tailor to the diverse needs of BVI people.

\section{Formative study}
We conducted a formative study to identify design requirements for a system providing live visual descriptions in the real world.
We conducted semi-structured interviews with 5 BVI participants (Table \ref{tab:demographic})
to gather feedback on their needs through several potential scenarios. 
We developed scenarios considering several aspects, including user intent, familiarity with environments, visual complexity, and sound contexts. 
We developed scenarios considering user intent, familiarity with environments, visual complexity, and sound contexts, as these aspects were identified in previous works as influencing information needs \cite{soundshift,stangl21assets,engel,kreiss2022context,stanglchi20,navcog,pathfinder}.
Participants were asked to imagine using a future live description system capable of capturing visuals and sounds in their surroundings and brainstorm their needs and potential solutions. From these discussions, we extracted key insights reflecting participants’ needs and strategies, which we used in the design of {\name}.

\subsection{Design Considerations}
We reported design considerations derived from our participants:

\textbf{D1 - Overview first, adaptive details on the fly.} 
Participants emphasized the need for descriptions with proper levels of granularity, depending on their context. They preferred immediate and succinct information when several important events occurred simultaneously (\textit{e.g.,} multiple barriers and directions during navigation), and longer and detailed descriptions when there was no time pressure (\textit{e.g.,} understanding artwork in a museum). When searching for something, they wanted an overview of the space, including landmarks and spatial locations, followed by more details as they approached the target or encountered similar items requiring differentiation. This approach aligns with the \textit{``Overview first, zoom and filter, then details-on-demand.''}  by Shneiderman \cite{shneiderman2003eyes}. Therefore, our solution should provide the proper level of information and delve into details when users express interest.

\textbf{D2 - Prioritize descriptions based on semantic relevance.} 
Participants mentioned strategies for filtering and prioritizing complex visual information. 
The most commonly noted strategy was to prioritize descriptions relevant to their goals of context, such as road signs or barriers during navigation, available stores and offerings during meal times, etc. 
They also emphasized that nearby objects are more important for safety and should be prioritized over distant information. Our system should present information most relevant to the user's goals and proximity to ensure timely and practical use.
    
\textbf{D3 - Enable customizability for varied user needs.} 
Similar to prior work in Section \ref{RW_varied}, we observed varied individual preferences from our participants.
They expressed different information needs depending on the context. For example, descriptions should consider their mobility, such as providing information about objects out of their cane reach (\textit{e.g.,} hanging lights, cars with high ground clearance) or focusing on dynamic obstacles but not static ones in their familiar environments. 
Participants also noted that sound context influences the consumption of descriptions, with some suggesting pausing or increasing the volume in noisy environments. 
Some preferred manual control over these options, while others pointed out that the automatic approach would benefit urgent or busy scenarios, such as navigation or if the description content is crucial.
Based on these findings, {\name} should offer customizable options for description content and presentation to meet diverse user needs.

\begin{figure*}[t!]
\vspace{-1pc}
\begin{center}
\includegraphics[width=\linewidth]{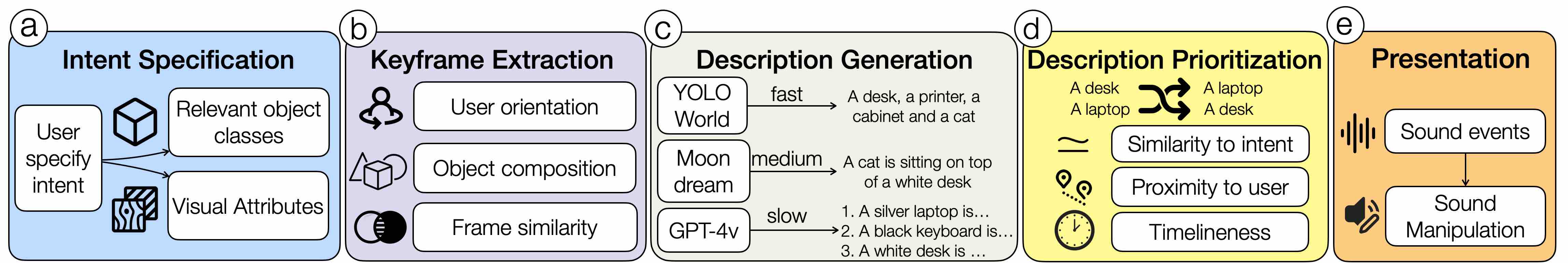}
\vspace{-1.5pc}
\caption{{\name} system architecture. 
(a) The user first specifies their intent through speech and {\name} decomposes it into specific visual attributes and relevant objects. 
(b) {\name} extracts keyframes based on user orientation, object compositions, and frame similarity.
(c) Next, it generates candidate descriptions with a suite of visual and language models. 
(d) {\name} then prioritizes the descriptions based on the user’s intent, proximity to the user, and relevance to the current visual context. 
(e) Finally, it detects environmental sounds and manipulates the presentation of the descriptions accordingly.
}
\vspace{-1pc}
\label{fig:system_diagram}
\Description{Figure 3
WorldScribe system architecture. 
Figure 3 a 
The user first specifies their intent through speech and WorldScribe decomposes it into specific visual attributes and relevant object classes. 
Figure 3 b 
WorldScribe extracts keyframes based on user orientation, object compositions, and frame similarity.
Figure 3 c
Next, it generates candidate descriptions with a suite of visual and language models. For instance, yolo world is the fastest one and generates description like “A desk, a printer, a cabinet and a cat”. Moondream is the second fast one and generates description like A cat is sitting on top of a white desk. GPT-4v is the slowest one and generates detailed object descriptions like 1. A silver laptop is … 2. A black keyboard is … 3. A white desk is ….
Figure 3 d 
WorldScribe then prioritizes the descriptions based on timeliness, richness, similarity to the user's intent and proximity to the user. 
Figure 3 e 
Finally, it detects environmental sounds and manipulates the presentation of the descriptions accordingly.
}
\end{center}
\end{figure*}

\section{{\name}}
{\name} is a system that provides live visual descriptions for BVI people to facilitate their environmental understanding. 
BVI users can specify their intent, general or specific, which will be decomposed by {\name} into specific visual attributes and objects of relevance (\textsc{Intent Specification Layer}). 
Then, {\name} extracts key frames from the camera video stream (\textsc{Keyframe Extraction Layer}), and employs a suite of vision and language models to generate rich descriptions (\textsc{Description Generation Layer}). 
The descriptions are prioritized based on users' intent, proximity, and timeliness (\textsc{Prioritization Layer}), and presented with audio manipulations based on sound context (\textsc{Presentation Layer}).

\begin{figure}[b]
\begin{center}
\vspace{-2pc}
\includegraphics[width=\linewidth]{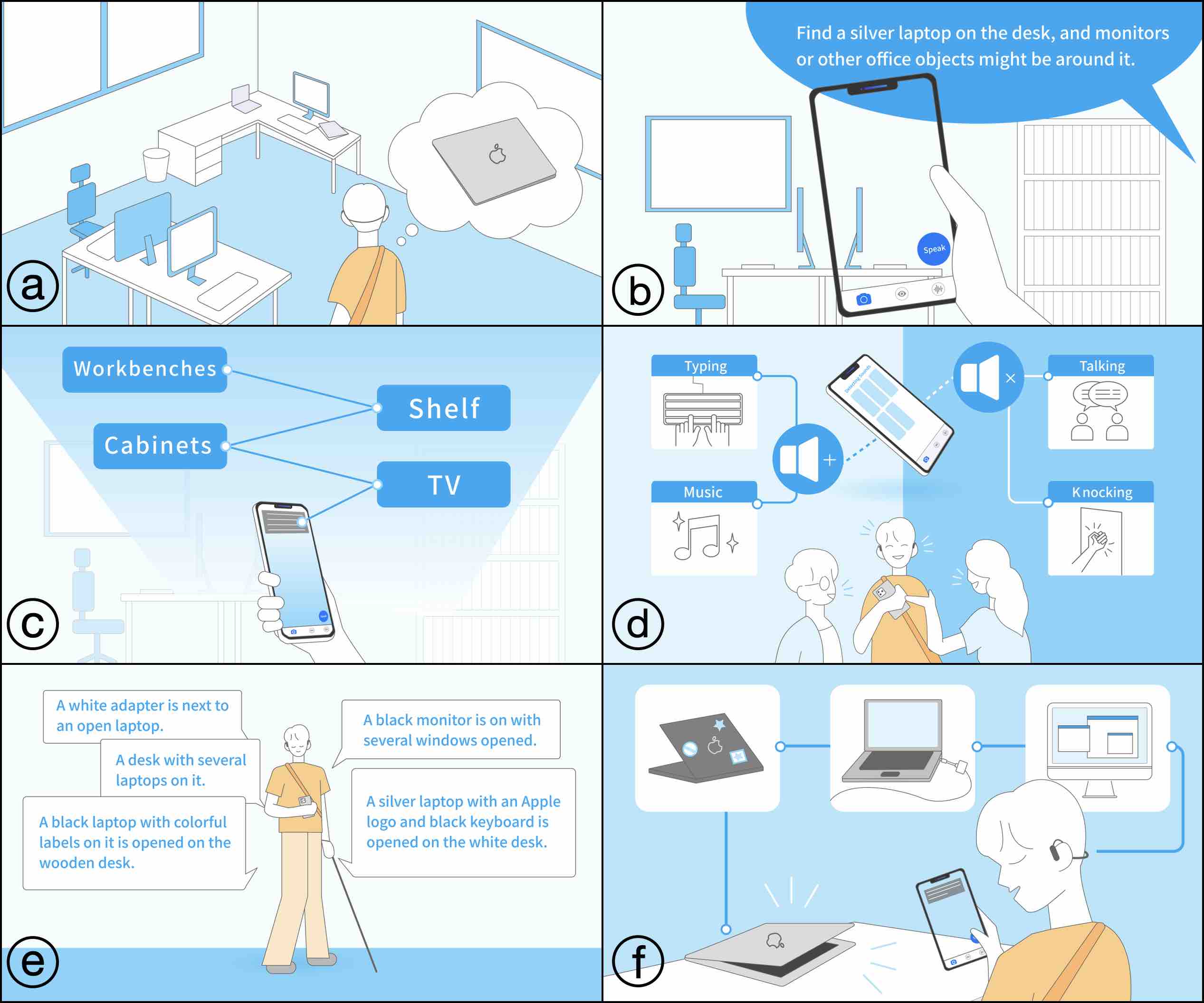}
\vspace{-1.5pc}
\caption{
(a) Brook is looking for a silver laptop using {\name} in the lab by first  
(b) specifying his intent. 
(c) As he moves quickly, {\name} reads out names of fixtures, and  
(d) pauses or increases its volume based on environmental sounds. When approaching his seat and Brook stops to scan, (e) {\name} provides verbose descriptions when the visual scene is relevant to his intent, (f) allowing him to follow the cues and find the laptop.
}
\label{fig:indoor_scene}
\Description{Figure 4
Figure 4a Brook is looking for a silver laptop using WorldScribe in the lab. The image show a man is walking in the space and a callout contains a silver laptop.
Figure 4b He is specifying his intent. The image shows a hand is holding the phone with WorldScribe interface opened and a callout contains the text “Find a silver laptop on the desk, and monitors or other office objects might be around it”. The visual scene has a lot of objects such as monitors, chairs, book shelves, and tables.
Figure 4c As he moves quickly,WorldScribe reads out names of fixtures. The image shows a hand holding the smartphone with WorldScribe opened and showing texts like Workbenches, shilf, TV, cabinets.  
Figure 4d WorldScribe pauses or increases its volume based on environmental sounds. The image shows that the man is talking to others and the phone will mute during people talking or knocking, and will increase the volume if someone is typing or playing music.
Figure 4e When approaching his seat and Brook stops to scan, WorldScribe provides verbose descriptions when the visual scene is relevant to his intent.  The image shows that the man is focusing on listening to the descriptions, such as “A white adapter is next to an open laptop”, “A desk with several laptops on it”, “A black laptop with colorful labels on it is opened on the wooden desk”, “A black monitor is on with several windows opened”, and “A silver laptop with an Apple logo and black keyboard is opened on the white desk”
Figure 4f This allows him to follow the cues and find the laptop. The image shows different objects such as monitor with content opened, desk top tethered with adaptors and a laptop with stickers on it, and finally linking to the silver laptop the man is looking for.
}
\end{center}
\end{figure}

\subsection{Scenario Walkthrough of {\name}}\label{walkthrough}
Here, we illustrate {\name} in an everyday scenario, taking Brook as the main character, a graduate student who is blind.

Brook just finished his advising meeting, and he wants to find a lab laptop with powerful computational resources to proceed with his project. The lab is filled with large items like TVs, workbenches with electronics, rows of seats with monitors, personal items, cabinets, and garbage bins, with their layout changing daily based on activities. The only cues Brook has from his labmate are that the laptop is silver (Figure \ref{fig:indoor_scene}a) and located around the student seats amid office or personal objects (\textit{e.g.,} monitors, adapters, backpacks).

Arriving at the spacious lab, Brook specifies his goal by talking to {\name} (Figure \ref{fig:indoor_scene}b): \textit{``find a silver laptop on the desk, and monitors or other office objects might be around it.''} He then aims his smartphone camera forward with {\name} on. Along the way, {\name} provides concise and timely descriptions of objects not directly related to his goal, where several fixtures, such as \PROMPT{TVs}, \PROMPT{cabinets}, \PROMPT{workbenchs}, help him orient himself (Figure \ref{fig:indoor_scene}c). 

As he approaches his seat, surrounded by relevant items like monitors, chargers, cables, and adapters, {\name} becomes more verbose (Figure \ref{fig:indoor_scene}e), providing descriptions, such as \PROMPT{A black monitor is on with several windows opened}, \PROMPT{A white adapter is next to an open laptop}, and \PROMPT{A desk with several laptops on it}. These help Brook ascertain that the laptop he seeks is nearby.

Brook then scans his surroundings slowly with {\name}, listening to the detailed descriptions with objects' visual attributes to help distinguish the laptop he is looking for (Figure \ref{fig:indoor_scene}f), such as \PROMPT{A black laptop with colorful labels on it is opened on the wooden desk}, and finally, \PROMPT{A silver laptop with an Apple logo and black keyboard is opened on the white desk}.

In the lab, people talk, type on keyboards, or use power tools, generating various noises. Brook is accustomed to these sounds and has customized {\name} to accommodate different interference (Figure \ref{fig:indoor_scene}d). For instance, when his labmates talk, Brook wants to join the conversation, so {\name} immediately pauses descriptions to let him listen and resumes when they stop talking. Also, if a cellphone or clock rings suddenly, {\name} automatically increases the description's volume to ensure he can hear it clearly.

\begin{figure}[t!]
\begin{center}
\includegraphics[width=\linewidth]{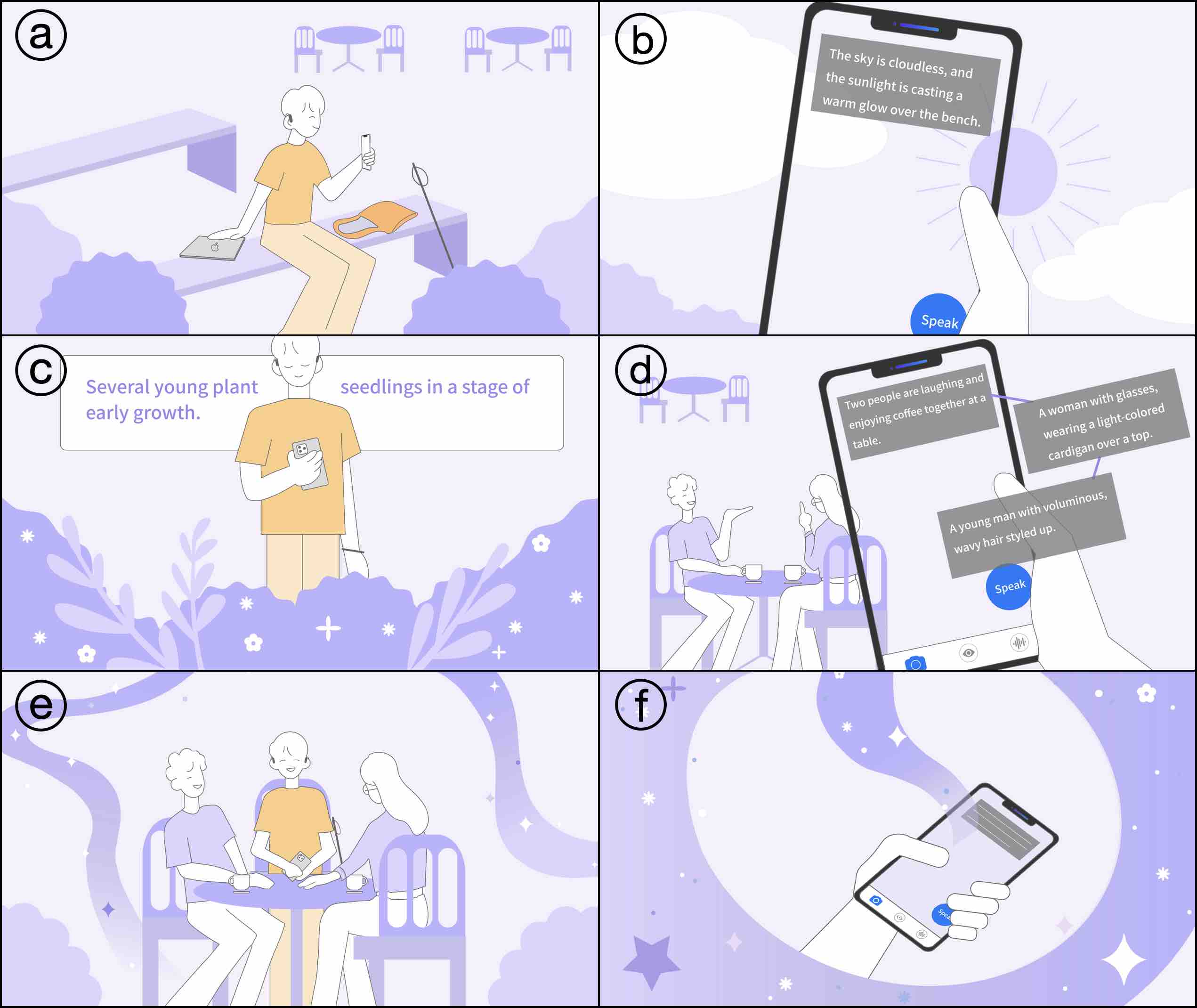}
\vspace{-1.5pc}
\caption{
(a) Brook takes a break on the balcony and uses {\name} to explore his surroundings.
(b) Through the live visual descriptions, he knows the sky is sunny, 
(c) plants are growing, and also notices 
(d) his friends are here. 
(e) He then joins them and has a delightful tea time. 
(f) {\name} facilitate the understanding and access of his surroundings, and make his day.
}
\vspace{-1pc}
\label{fig:outdoor_scene}
\Description{Figure 5
Figure 5a Brook takes a break on the balcony and uses WorldScribe to explore his surroundings. The image shows a man is sitting on the bench with his cane, silver laptop and a bag on the side.
Figure 5b Through the live visual descriptions, he knows the sky is sunny. The image shows WorldScribe app is being used to capture the sky.
Figure 5c The image shows there are leaves and flowers are glowing. 
Figure 5d The image shows a hand is holding a phone using WorldScribe to capture two people who are sitting at a table and chatting.
Figure 5e He then joins them and has a delightful tea time. The image shows the man is joining in the conversation and sitting the same table with the two people.
Figure 5f WorldScribe facilitate the understanding and access of his surroundings, and make his day. The image shows some spiral visual effects comes out of the phone with WorldScribe opened.
}
\end{center}
\end{figure}

After working for a while, Brook takes a break on the building's balcony and uses {\name} to explore his surroundings (Figure \ref{fig:outdoor_scene}a). The balcony has several plants, benches on the sides, and coffee tables. When Brook aims the camera at the sky (Figure \ref{fig:outdoor_scene}b), {\name} describes \PROMPT{The sky is cloudless, and the sunlight is casting a warm glow over the bench}. Brook then turns to the plants (Figure \ref{fig:outdoor_scene}c), wondering if they have begun to germinate in this early spring period, and {\name} describes: \PROMPT{Several young plant seedlings in a stage of early growth.}
The beautiful view revitalizes his weary mind from work. Later, Brook recognizes familiar voices coming from a coffee table. Turning towards the sound (Figure \ref{fig:outdoor_scene}d), {\name} describes: \PROMPT{Two people are laughing and enjoying coffee together at a table}. It then provides detailed descriptions such as \PROMPT{A woman with glasses, wearing a light-colored cardigan over a top} and \PROMPT{A young man with voluminous, wavy hair styled up}. With this detailed visual information and the voices, Brook identifies them, joins their conversation, and enjoys a delightful tea time with them (Figure \ref{fig:outdoor_scene}e \& f).

\subsection{User Interface}\label{mobile_interface}
\begin{figure}[t!]
\begin{center}
\includegraphics[width=\linewidth]{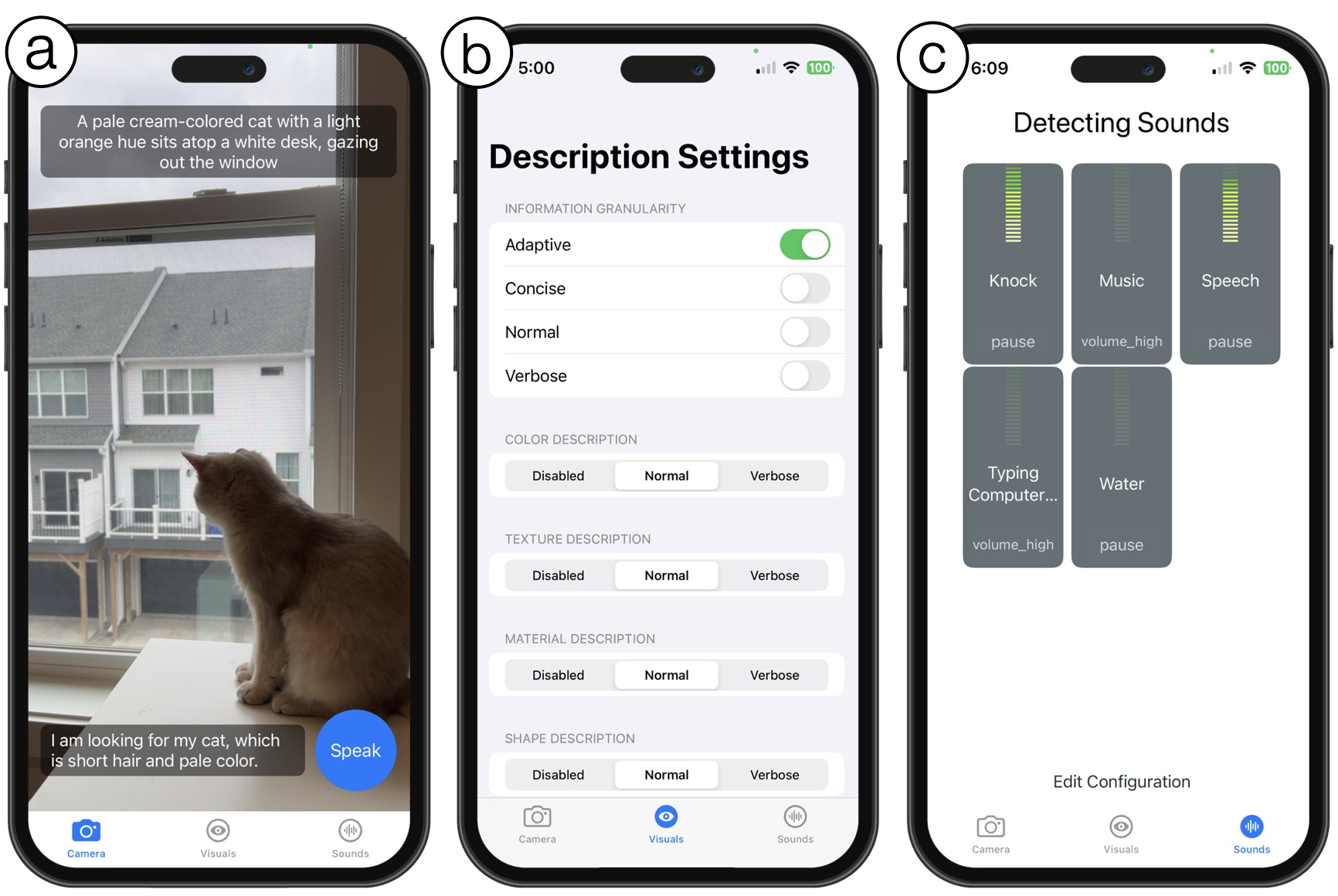}
\vspace{-1.5pc}
\caption{
{\name} user interface. 
(a) The user can specify their intent and needs regarding visual attributes or audio presentation through speech input. (b) Besides speech, they can manually select options for richness and other visual attributes.
(c) They can also configure pauses or increase the volume of descriptions if certain sound events are detected.
}
\vspace{-1pc}
\label{fig:app}
\Description{Figure 6
WorldScribe user interface. 
Figure 6 a 
The user can specify their intent and needs regarding visual attributes or audio presentation through speech input. The image show a screenshot of the app. There is a description on the top “A pale cream-colored cat with a light orange hue sits atop a white desk, gazing out the window.” The is a sentence on the bottom showing user’s intent “I am looking for my cat, which is short hair and pale color.” And there is a “Speak” button besides the intent description. There are three tabs in a row, which are camera, visuals, and sounds. 
Figure 6 b 
Besides speech, they can manually select options for richness and other visual attributes. This is the screenshot of the description settings. The user can customize information granularity with levels like adaptive, concise, normal, and verbose. The user can also customize visual features such as color description, texture description, material description, shape description with three levels including disabled, normal and verbose.
Figure 6 c
They can also configure pauses or increase the volume of descriptions if certain sound events are detected. This is the screenshot of the sound customization setting. The image shows some settings such as pause for knock, volume_high for music, pause for Speech, Volume_high for typing computer and pause for water.
}
\end{center}
\end{figure}

{\name} has a mobile user interface that takes the user's camera stream, environmental sounds, and user customizations as inputs for generating descriptions (Figure \ref{fig:app}). 
The interface includes three pages: \textit{(i)} a main page with a camera streaming view and speech interface, \textit{(ii)} a customization page for visual information, and \textit{(iii)} a page for customizing audio presentation.
Users can open the camera on the main page and use speech to indicate their intent (Figure \ref{fig:app}a), such as \PROMPT{find my cat, short hair and pale brown.}
To customize visual attributes of their interest, users can also use the speech on the main page, such as \PROMPT{I am interested in color} or \PROMPT{Tell me everything is pale brown}, or manually toggle options on or off (Figure \ref{fig:app}b).
Similarly, users can verbally change the presentation of descriptions, such as \PROMPT{Pause when someone talks} or \PROMPT{Increase volume during ringtone}, or manually select the options through the picker (Figure \ref{fig:app}c).

\subsection{Intent Specification Layer}
\label{customization_layer}
\begin{figure}[b!]
\vspace{-1pc}
\begin{center}
\includegraphics[width=\linewidth]{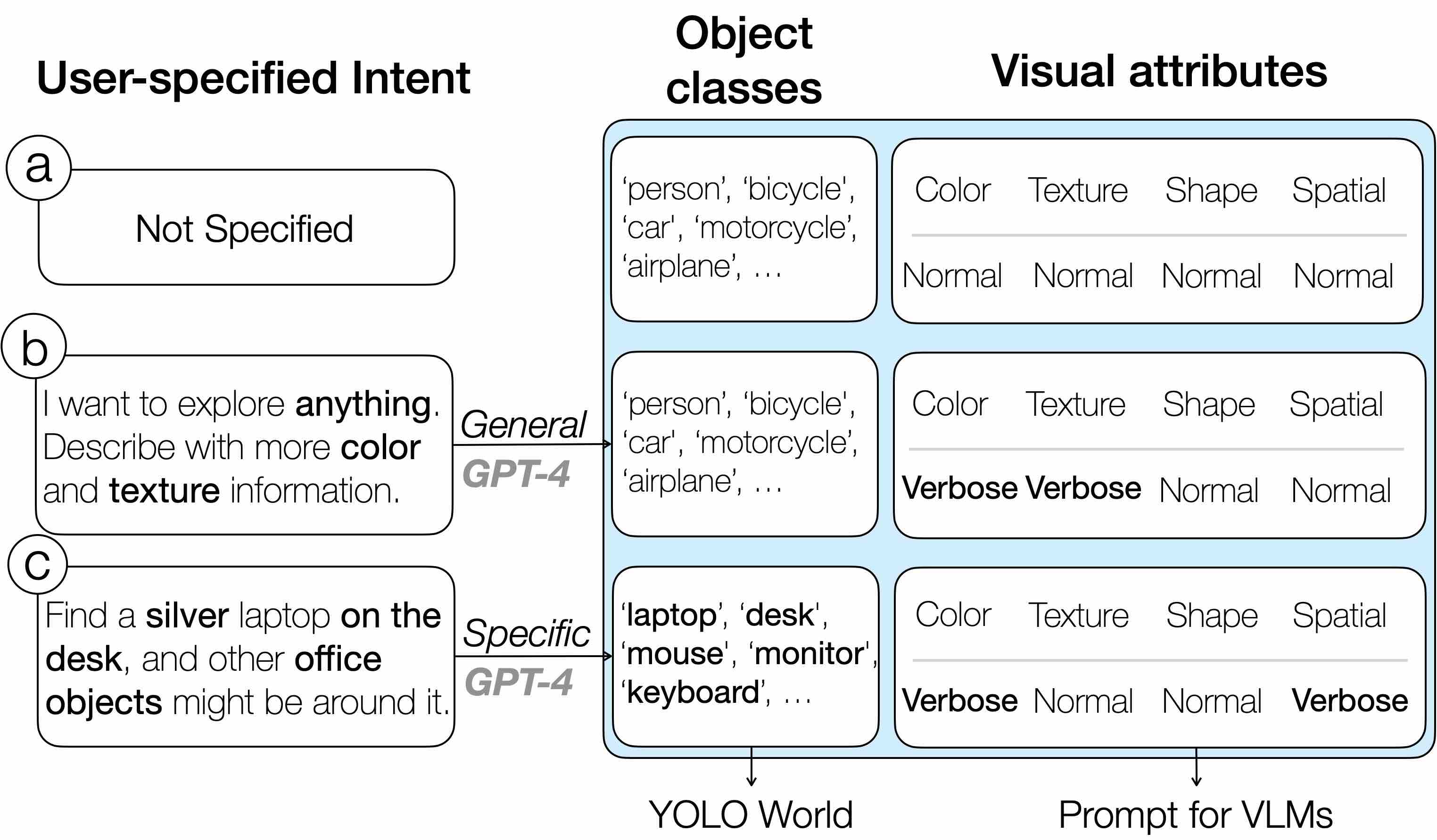}
\vspace{-2pc}
\caption{{\name} classifies the user's intent into \textit{general} or \textit{specific}, and generates relevant object classes and visual attributes by prompting GPT-4 \cite{gpt4}. 
(a) By default, {\name} uses classes from established datasets (\textit{e.g.,} COCO \cite{lin2014microsoft}, Object365 \cite{shao2019objects365}), and sets visual attributes to \textit{Normal}.
(b) If the intent is classified as \textit{general}, {\name} adjusts the visual attributes of interest to \textit{Verbose}.
(c) If the intent is classified as \textit{specific}, {\name} generates relevant object classes and sets visual attributes to \textit{Verbose}.
}
\vspace{-0.5pc}
\label{fig:intent}
\Description{Figure 7
WorldScribe decomposes the user's intent into general or specific, and generates relevant object classes and visual attributes by prompting GPT-4. 
Figure 7 a 
By default, WorldScribe uses classes from established datasets (e.g., COCO), and sets visual attributes to Normal. Object classes include person, bicycle, car, motorcycle, airplane, etc. The visual attributes include color-normal, texture-normal, shape-normal, spatial-normal.
Figure 7 b 
If the intent is classified as general, WorldScribe further tunes the visual attributes of the user's interest to Verbose. The intent is “I want to explore anything. Describe with more color and texture information”. Object classes include person, bicycle, car, motorcycle, airplane, etc. The visual attributes include color-verbose, texture-verbose, shape-normal, spatial-normal.
Figure 7 c 
If the intent is classified as specific, WorldScribe generates relevant object classes and tunes visual attributes to Verbose. The intent is “find a silver laptop on the desk, and other office objects might be around it”. Object classes include laptop, desk, mouse, monitor, keyboard. The visual attributes include color-verbose, texture-normal, shape-normal, spatial-verbose.
}
\end{center}
\end{figure}

In this layer, {\name} aims to obtain the user's intent and needs on visual information to enable customizability (\textbf{D3}). 
Users specify their intent on the mobile interface, and {\name} will classify it as \textit{general} or \textit{specific} and generate relevant object classes and visual attributes by prompting GPT-4 \cite{gpt4}. 
If the intent is \textit{general} or \textit{not specified} (Figure \ref{fig:intent}a\&b), {\name} takes object classes from existing datasets (\textit{e.g.,} COCO \cite{lin2014microsoft}, Object365 \cite{shao2019objects365}).
If the intent is \textit{specific} (\textit{e.g.,} \PROMPT{Find a silver laptop on the desk, and other office objects might be around it}), {\name} prompts GPT-4 \cite{gpt4} to generate a list of relevant objects (\textit{e.g.,} \PROMPT{[laptop, desk, monitor, ...]}) and adjust visual attributes of interest (\textit{e.g.,} \textit{color} and \textit{spatial information}) to \textit{Verbose}.
{\name} then uses YOLO World \cite{yolow} and ByteTrack \cite{bytetrack} to support open vocabulary object recognition and tracking. 
Users can further refine the generated object classes and other visual attributes through speech (Figure \ref{fig:app}a) or manually on the customization page (Figure \ref{fig:app}b).

\begin{figure*}[t]
\vspace{-1pc}
\begin{center}
\includegraphics[width=\linewidth]{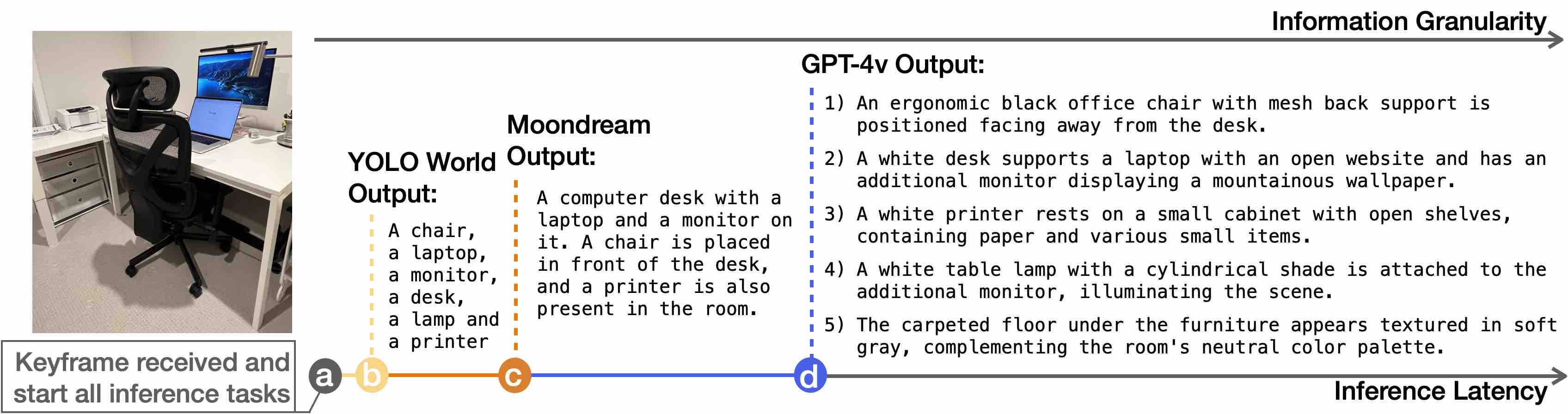}
\vspace{-1.8pc}
\caption{
{\name} description generation pipeline with different inference latency and granularity.
(a) Upon receiving a keyframe, {\name} starts all visual description tasks.
(b) First, YOLO World \cite{yolow} identifies objects as word-level labels in real-time (.1s).
(c) Second, Moondream \cite{moondream} generates short descriptions with objects and spatial relationships, with a small delay (\textasciitilde3s).
(d) Finally, GPT-4v \cite{gpt4v} provides detailed descriptions with visual attributes, with a longer delay (\textasciitilde9s).
The estimated inference time in each model was calculated based on our computing platforms and log data in our user evaluation. 
}
\vspace{-0.5pc}
\label{fig:granularity}
\Description{Figure 8 
WorldScribe description generation pipeline with different inference latency and granularity.
Figure 8 a 
Upon receiving a keyframe, WorldScribe starts all visual description tasks.
Figure 8 b 
First, YOLO World identifies objects as word-level labels in real-time (.1s). The image shows texts “Yolo World Output: A chair, a laptop, a monitor, a desk, a lamp, and a printer”
Figure 8 c 
Second, Moondream generates short descriptions with objects and spatial relationships, with a small delay (~3s). The image shows texts “Moondream Output: A computer desk with a laptop and a monitor on it. A chair is placed in front of the desk, and a printer is also present in the room.”
Figure 8 d 
Finally, GPT-4v provides detailed descriptions with visual attributes, with a longer delay (~9s).
The estimated inference time in each model was calculated based on our computing platforms and log data in our user evaluation. The image shows texts “GPT-4v Output: (1) An ergonomic black office chair with mesh back support is positioned facing away from the desk. (2) A white desk supports a laptop with an open website and has an additional monitor displaying a mountainous wallpaper. (3) A white printer rests on a small cabinet with open shelves, containing paper and various small items. (4) A white table lamp with a cylindrical shade is attached to the additional monitor, illuminating the scene. (5)The carpeted floor under the furniture appears textured in soft gray, complementing the room's neutral color palette.”
}
\end{center}
\end{figure*}

\subsection{Keyframe Extraction Layer}\label{keyframe_layer}
In this layer, {\name} aims to identify keyframes that indicate salient visual changes or user interests in the visual scene.
To achieve this, our approach uses two methods: camera orientation and visual similarity.
First, {\name} monitors changes in the camera's orientation using the phone's inertial measurement unit (IMU). A keyframe is selected whenever the camera's orientation shifts by at least 30 degrees (one clock unit) from the previous keyframe, indicating a possible turn into a new visual scene.

Second, {\name} determines a keyframe by analyzing visual changes across frames.
To minimize detection errors, such as misassigned object classes or IDs, we assess the consistency of object composition over $n$ consecutive frames.
In each $i^{\text{th}}$ frame, a detected object is represented as $(ID_i, C_i)$, where $ID_i$ is the object's index, and $C_i$ is the class. Thus, all objects in the $i^{\text{th}}$ frame are represented as $O_i = {(ID_i, C_i)}$.
A keyframe is identified if the object composition remains consistent across $n$ frames, denoted as $O_i = O_{i+1} = ... = O_{i+n-1} \not = \emptyset$. The $(i+n-1)^{\text{th}}$ frame is then taken as the keyframe.
Furthermore, to determine if the user is interested in a visual scene and requires details (conforming to \textbf{D1}), we check the $m$ consecutive keyframes with the same composition and use the latest keyframe to prompt VLMs for detailed descriptions. 

In scenarios where the object compositions across $n$ consecutive frames are empty, denoted by $O_i = O_{i+1} = ... = O_{i+n-1} = \emptyset$, it suggests that the predefined object classes may not cover the objects in the scene, resulting in false negatives. Therefore, we still take the $(i+n-1)^{\text{th}}$ frame as the keyframe.
To eliminate genuinely empty scenes (\textit{e.g.,} aiming at a plain white wall), we measure the similarity between the candidate frame and the previous keyframe. 
We calculate the cosine similarity ($\textit{cos\_sim}$) between the image feature vectors of the two frames, extracted from the FC2 layer of the VGG16 \cite{vgg16}. If $\textit{cos\_sim}$ is lower than a threshold $thres$, we count the frame as a keyframe.
Furthermore, we observed situations where object compositions differ across consecutive $n$ frames, denoted by $O_i \not = O_{i+1} \not = ... \not = O_{i+n-1} \not = \emptyset$. These changes often indicate camera drifting or objects moving in and out of view. In such cases, we check every $2n$ frame, selecting the $(i+2n-1)^{\text{th}}$ frame as a keyframe if the condition is satisfied. 
In our implementation, we empirically set $n=5$, $m=3$, and $thres=0.6$.

\subsection{Description Generation Layer}\label{generation_pipeline}
In this layer, {\name} aims to generate descriptions with adaptive details to the user's intent and visual contexts.
To achieve this, {\name} leverages a suite of VLMs that balances the tradeoffs between their richness and latency to support real-time usage (Figure \ref{fig:granularity}).

To provide overview first and details on the fly (\textbf{D1}), {\name} recognizes objects by YOLO World \cite{yolow} and structures its results into short phrases, \textit{e.g.,} \PROMPT{A chair, a laptop, a monitor, ...}, allowing it to provide an overview of objects in the visual scene in real time (Figure \ref{fig:granularity}b). 
Then, {\name} describes objects and their spatial relationship by prompting Moondream \cite{moondream} (Figure \ref{fig:granularity}c), a compact vision language model, that can achieve a decent performance in terms of latency and accuracy on this information based on our observation.
Finally, {\name} prompts GPT-4v \cite{gpt4v} to generate descriptions of different levels of details based on user contexts (\textbf{D1}).
It offers three levels of detail: \textit{Verbose}, \textit{Normal}, and \textit{Concise}, associated with prompts specifying \textit{e.g.,} \PROMPT{over 15 words}, \PROMPT{at least 10 words}, and \PROMPT{less than 5 words}, respectively. {\name} dynamically adjusts these length constraints based on visual complexity.
For instance, it becomes \textit{Verbose} if the visual scene is focused on for a long period, indicating user interest, with consecutive keyframes identified (See Section \ref{keyframe_layer}). 
It becomes \textit{Concise} when multiple objects of interest are detected in consecutive keyframes to ensure timely coverage; otherwise, it remains \textit{Normal} by default.
Along with the recognized visual attributes of interest, {\name} dynamically creates prompts to suit users' intent (\textbf{D3}):
\begin{quote}
``You are a helpful visual describer, who can see and describe for BVI people. You will not mention this is an image; just describe it, and also don't mention camera blur or motion. Please ensure you provide these adjectives to enrich the descriptions \emph{[desired visual attributes (\textit{e.g.,} color, texture, shape, spatial), with examples adjectives]}, you should describe each object with ONLY ONE sentence at maximum. Don't use 'it' to refer to an object. Most importantly, each sentence should be \emph{[sentence length constraint \textit{e.g.,  verbose, normal or concise}]}.''
\end{quote}

Each short phrase from YOLO World \cite{yolow}, general description from Moondream \cite{moondream}, and detailed description from GPT-4v \cite{gpt4v} are stored as a packet in the description buffer, and {\name} selects the most relevant and up-to-date one based on the user's contexts, which we describe next.

\begin{figure}[b]
\begin{center}
\vspace{-1.5pc}
\includegraphics[width=\linewidth]{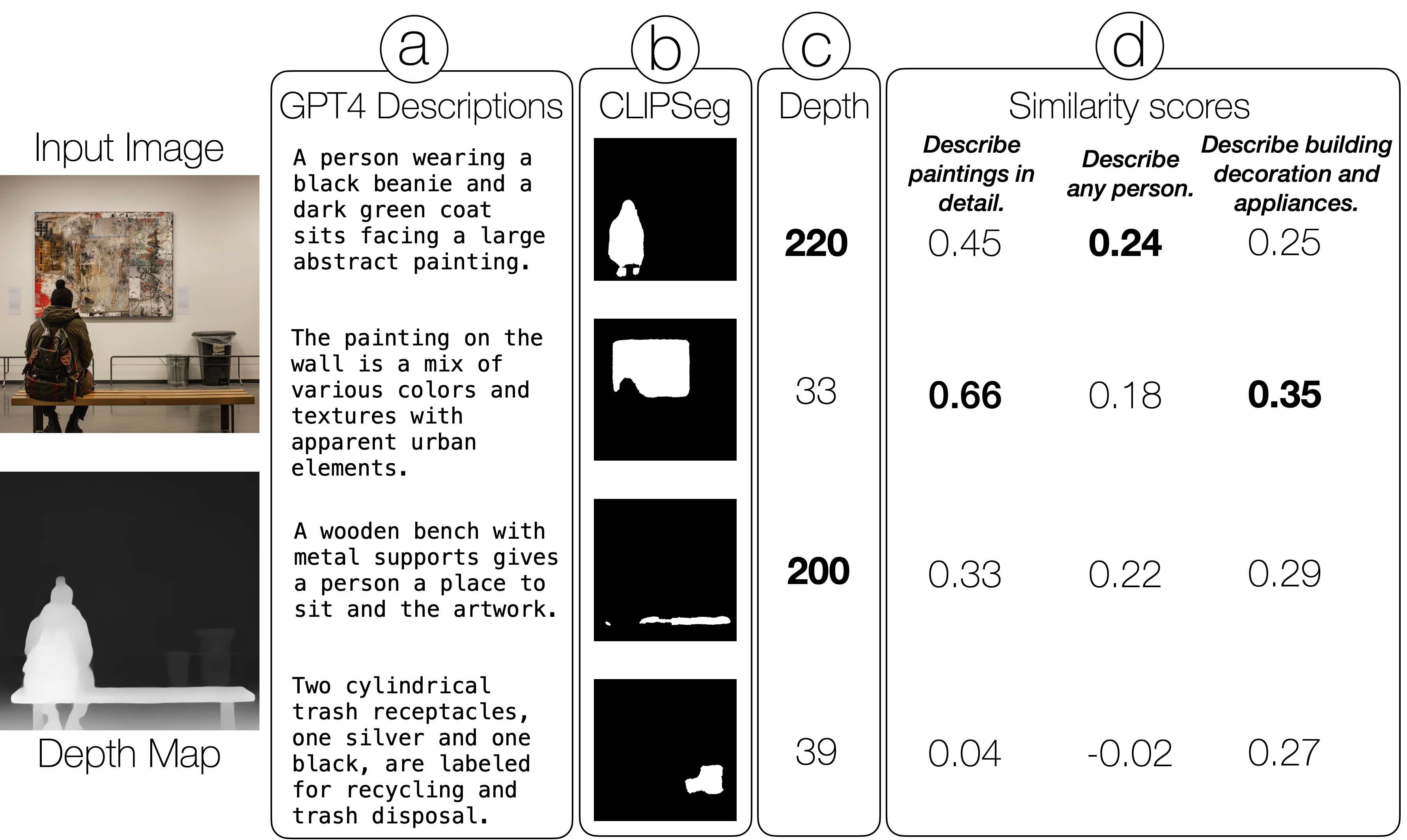}
\vspace{-1.5pc}
\caption{{\name} pipeline to prioritize descriptions based on semantic information. 
(a) Given an input keyframe image, GPT-4v \cite{gpt4v} generates descriptions on individual objects.
(b) CLIPSeg \cite{clipseg} generates the cropped image content based on each description.
(c) The average depth value of the CLIPSeg-cropped area was computed for each described content with the depth map generated by Depth Anything \cite{yang2024depth}; the higher the depth map value, the nearer to the user.
(d) The similarity score was also computed between each description and the user's specified intent. 
Finally, {\name} prioritize GPT-4v \cite{gpt4v} generated descriptions based on semantic relevance to the user's intent, then proximity to the user (\textasciitilde1s).
}
\vspace{-1pc}
\label{fig:priority}
\Description{Figure 9
WorldScribe pipeline to prioritize descriptions based on semantic information. 
Figure 9 a 
Given an input keyframe image, GPT-4v generates descriptions on individual objects. This column show four descriptions in separate rows, from top to bottom: “A person wearing a black beanie and a dark green coat sits facing a large abstract painting.”, “The painting on the wall is a mix of various colors and textures with apparent urban elements.”, “A wooden bench with metal supports gives a person a place to sit and the artwork.”, “Two cylindrical trash receptacles, one silver and one black, are labeled for recycling and trash disposal.”. 
Figure 9 b 
CLIPSeg generates the cropped image content based on each description. The images also shows four separate rwos indicating the depth segmentation maps for each object.
Figure 9 c 
The average depth value of the CLIPSeg-cropped area was computed for each described content with the depth map generated by Depth Anything; the higher the depth map value, the nearer to the user. Depth scores from top to bottom: 220 (bold text), 33, 200 (bold text), 39
Figure 9 d 
The similarity score was also computed between each description and the user's specified intent. 
Finally, WorldScribe prioritize GPT-4v generated descriptions based on semantic relevance to the user's intent, then proximity to the user (~1s). The images show the similarity scores based on descriptions “Describe paintings in detail” with scores from top to bottom 0.45, 0.66 (bold text), 0.33, 0.04, “Describe any person” with scores from top to bottom: 0.24 (bold text), 0.18, 0.22, -0.02, and “Describe building decoration and appliances” with scores from top to bottom: 0.25, 0.35 (bold text),0.29, 0.27
}
\end{center}
\end{figure}

\subsection{Description Prioritization Layer}\label{prioritization_layer}
In this layer, {\name} aims to select a description based on the match to the user's intent, proximity to the user, and relevance to the current visual context (\textbf{D2}).

\subsubsection{Sorting GPT-4v detailed descriptions by semantic relevance}
Descriptions from YOLO World \cite{yolow} and Moondream \cite{moondream} provide an overview and general information, which should always be prioritized to help the user construct an initial understanding of the new visual scene (\textbf{D1}).
In contrast, detailed descriptions from GPT-4v \cite{gpt4v} may contain information irrelevant to the user's intent. 
Therefore, {\name} ranks the descriptions generated by GPT-4v \cite{gpt4v} based on their relevance to the user's intent and the proximity of the described content to the user; the nearer, the earlier to describe.

To achieve this, we first get a set of descriptions from GPT-4v \cite{gpt4v} $S = \{s_i\}$.
For a description $s_i$, we compute the sentence similarity score $\text{SIM}(s_i)$ to the user's intent, and a depth score $Depth(s_i)$.
To calculate the depth score, we extract the subject and its descriptors from a description (\textit{e.g.,} \PROMPT{Two cylindrical trash receptacles} in Figure \ref{fig:priority}), locate each in the frame using CLIPSeg \cite{clipseg} (Figure \ref{fig:priority}b), and crop the salient area on the depth map generated by Depth Anything \cite{yang2024depth}. 
We then compute the average depth $Depth(s_i)$ of the cropped area for each description $s_i$. 

To sort the descriptions in $S$, We divide them into two sets $S_a = {s_i|SIM(s_i)>=\text{Threshold}}$, and $S_b = {s_j|SIM(s_j)<\text{Threshold}}$. 
We then sort the descriptions in $S_a$ as a sorted sequence $A = (s_0, s_1, ..., s_{n-1})$ based on the similarity score such that $SIM(s_i)>=SIM(s_{i+1})$. 
We also sort sentence in $S_b$ as $B = (s_n, s_{n+1}, ... s_{n+m-1})$ based on the depth score such that $Depth(s_j)>=Depth(s_{j+1})$. 
Finally, we concatenate the two sequences into final one, $A^\frown B = (s_0, ..., s_{n-1}, s_n, ..., s_{n+m-1})$. 
This approach ensures the initial descriptions are highly relevant to the user's intent, regardless of proximity. 
The remaining descriptions are sequenced by their proximity, with nearer elements described sooner.

\subsubsection{Selecting up-to-date description based on current context}\label{select_new_description}
In reality, users may move or turn frequently, and the environment can change significantly, leading to frequent visual changes and generating multiple keyframes and VLM requests in a short time. 
As VLM inference times vary, some descriptions may become outdated if they take too long to process and will not be useful to the user's current visual context. Thus, we need to select descriptions that best represent the current visual scene.
We consider four criteria: camera orientation, object compositions, frame similarity, and similarity to previous descriptions. A description is selected if it satisfies any of these criteria.
First, we check if the candidate description's referenced object composition matches the current scene. Second, we compare the user's orientation from the description's referenced frame to the current orientation. Third, we compare the frame similarity between the description's referenced frame and the current frame using feature vectors extracted through VGG16 \cite{vgg16}. 
Descriptions similar to preceding spoken descriptions are skipped, and the description buffer is renewed when a description is omitted. Descriptions generated by GPT-4v \cite{gpt4v} are prioritized, followed by those from Moondream \cite{moondream} and YOLO World \cite{yolow}.

\subsection{Presentation Layer}\label{audio_presentation}
In this layer, {\name} aims to make descriptions audibly perceivable to the user by considering users' sound context.
To achieve this, {\name} runs a sound detection module in the background and automatically manipulates the presentation of the descriptions accordingly.
Based on our formative study, {\name} enables two audio manipulations on descriptions for the user to better perceive the description content in the noisy environment:
\textit{(i) Pausing} and \textit{(ii) Increasing volume}.  
Users can find sound events that interest them and customize the corresponding manipulations on descriptions in {\name} app (Figure \ref{fig:app}c).

\begin{table*}[t!]
  \caption{Participant demographics information. Participants in our formative study were marked as F1-F5. Participants in our user evaluation were marked as P1-P6.}
  \label{tab:demographic}
  \vspace{-1pc}
  \begin{center}
  \begin{tabular}{|l|l|l|p{5cm}|p{3cm}|p{5.5cm}|}
    \hline
    \textbf{ID} & \textbf{Age} & \textbf{Gender} & \textbf{Self-Reported Visual Ability} & \textbf{Assistive App Use} & \textbf{Self-Defined Goal in User Evaluation} \\
    \hline
    F1 & 34 & Male & Blind, since birth. Light perception. & BeMyEyes & N/A \\
    \hline
    F2 & 25 & Male & Blind, later in life. Light perception. & BeMyEyes, ENVision, TaptapSee, VoiceVISTA & N/A \\
    \hline
    F3 & 23 & Female & Blind, later in life. Light perception. & None & N/A \\
    \hline
    F4 & 35 & Male & Blind, later in life. Light perception. & BeMyEyes and Aira & N/A \\
    \hline
    F5 & 24 & Male & Blind, since birth. Light perception. & BeMyEyes & N/A \\
    \hline \hline
    P1 & 62 & Male & Low Vision, can't pick up details, using magnifiers. & None & Describe things on the wall. \\
    \hline
    P2 & 53 & Female & Blind, since birth. & SeeingAI, BeMyEyes & Describe posters and people in detail.\\
    \hline
    P3 & 60 & Female & Low Vision, can't pick up details. & None & Describe paintings or pictures in detail. \\
    \hline
    P4 & 40 & Male & Blind, since birth. Light perception. & SeeingAI, BeMyEyes and SoundScape & Describe any person.\\
    \hline
    P5 & 87 & Female & Low Vision, can't pick up details. & None & Describe artworks or paintings. \\
    \hline
    P6 & 72 & Female & Blind, since birth. Light perception. & SeeingAI, BeMyEyes, BeMyAI, Aira & Describe things in general with more color and texture information.\\
    \hline

  \end{tabular}
  \end{center}
  \Description{}
  \vspace{-1pc}
\end{table*}

\subsection{Implementation Details}\label{implementation_detail}
{\name} servers included a local server running on a Macbook M1 Max and another remote server with two embedded Nvidia GeForce RTX 4090. 
{\name} mobile app was built on an iPhone 12 Pro and streamed the camera frames to the local server through a Socket connection. 
YOLO World \cite{yolow} and ByteTrack \cite{bytetrack} were run on the local server with 5 frames per second (FPS) along with other algorithms, while other models in description generation and prioritization pipeline, including Moondream \cite{moondream}, Depth Anything \cite{yang2024depth} and CLIPSeg \cite{clipseg} were run on the remote server for each keyframe, as well as the pre-trained model \PROMPT{all-MiniLM-L6-v2} for sentence similarity from an open-sourced implementation on Huggingface. 
Overall, based on the data collected in our user study (Section \ref{user_evaluation}), {\name} achieved an overall latency of 1.44s, and each component took an average: YOLO World 0.1s, Moondream 2.87s, GPT-4v 8.78s, and prioritization pipeline 0.83s. 
For sound recognition, we used Apple's Sound Analysis example repository \cite{applesound}, which provides a visualization interface (Figure \ref{fig:app}c) and can identify over 300 sounds.

\section{User Evaluation}\label{user_evaluation}
We conducted a user evaluation with six BVI participants, where they used {\name} in three different contexts. 
This study aimed to explore 
\textbf{RQ1:} How do users perceive {\name} descriptions in various contexts?
and \textbf{RQ2:} What are the gaps between {\name} descriptions and users' expectations?
We detail our study method and results below.

\subsection{Participants}
We recruited six BVI participants (2 Male and 4 Female) through public recruitment posts on local associations of the blind. 
Participants aged from 40 to 87 (Avg. 62.3) and described their visual impairment as blind (N=3) or having residual vision (N=3).
Some participants had prior experiences using RSA services and used AI-enabled services, such as BeMyEyes \cite{bemyeyes} or SeeingAI \cite{seeingai} in their daily lives (Table \ref{tab:demographic}).

\subsection{Study Sessions}
We enacted three different scenarios: 
\textit{(i)} specific intent,
\textit{(ii)} general intent, and
\textit{(iii)} user-defined intent.
In each session, the descriptions were automatically paused if the speech was detected, including the conversation between participants and the experimenter, and the volume was automatically increased if a ringtone occurred.

\textbf{Scenario with specific intent.}
The first scenario, similar to the walkthrough scenario (Section \ref{walkthrough}), happened in our lab space, which is furnished with glass walls, wall-mounted TVs, several work benches with electronics and equipment, several rows of seats with monitors and scattered personal items, and a small kitchen area with microwave, fridge, sink and a lot of cabinets and garbage bins at the corners. 
The user's intent is \PROMPT{find a silver laptop on the desk, and monitors or other office objects might be around it.}
This scenario was designed to encourage them to think about the descriptions they need for specific purposes and whether {\name} supplements or obscures their intent.

\textbf{Scenario with general intent.}
This scenario happened on one of our building's floors, which has many common objects on the intricate hallways, such as poster stands, carts for construction, trash cans, desks, and sofas. 
On the wall or doors, there were several artworks, paintings, posters or emergency plans, and TVs. 
Random people were also walking in the hallway or meeting at public tables during the study. 
The intent of the scenario is \PROMPT{I am exploring a school building. Describe general information on the appliances and the building decorations.} 
This scenario was designed to prompt users to think if {\name} descriptions support their understanding of the environment. 

\textbf{User-defined scenario.}
After experiencing the previous scenarios, participants were asked to develop their own defined scenarios. 
They can also customize their desired visual attributes in {\name} mobile app based on their needs. 
We then took participants to the place they wanted to explore near our experiment sites. 

\textbf{Limitations.}
Though we tried to create different real-world scenarios, our study was conducted within our local environment and buildings.
This setting may not fully capture the diversity and complexity of real-world environments, potentially limiting the generalizability of our findings to other contexts.

\subsection{Procedure}
After providing informed consent, participants were introduced to {\name} and the functionalities they could customize, and experienced through each session. 
Participants opted to either hold the camera on their front or wear the lanyard smartphone mount we prepared. 
To facilitate the study progress and avoid fatigue, we kept each exploration for around ten minutes or until participants paused spontaneously.
At the end of each session, we interviewed our participants about their experiences with {\name}.
The study took about two hours, and participants were compensated \$50 for their participation. This study was approved by IRB in our institution.

\subsection{Measures and Analysis}
We asked our participants to comment on their perceived accuracy and quality of descriptions, their confidence in {\name} descriptions, and several other open-ended questions.
We recorded and transcribed the interviews and recorded all interactions with {\name}, which was also used for our pipeline evaluation (Section \ref{technical_evaluation}). 
Two researchers coded all qualitative interview feedback received in all sessions for further analysis via affinity diagramming.

\subsection{Results}

\subsubsection{Perceived accuracy and skepticism towards the descriptions}\label{accuracy}
\textbf{Participants perceived {\name} descriptions as accurate based on the contextual clues they ascertained but remained skeptical due to a few observed erroneous instances.
}
Participants generally commended {\name} for providing information otherwise unavailable in their everyday lives. 
They appreciated its constant descriptive capabilities, finding them useful for daily tasks such as grocery shopping, locating dropped items, and exploring the outdoors.
Participants valued the real-time feedback and considered the descriptions accurate and responsive. For instance, some (P2, P4) tested the system by placing their hands in front of the camera and received immediate descriptions of their hands and accessories:
\begin{quote}
    \textit{``I just wanted to test if it can describe the rings on my hand, it's like wow it did describe, and did a pretty good job and so responsive, so I think it's accurate for what it sees.''} - P2
\end{quote}

Despite acknowledging the accuracy and timeliness of {\name}'s descriptions, participants expressed tentative skepticism about its practical use due to several factors. For example, occasional hallucinations, such as detecting motorcycles in the building lobby or bikes in the office space, impacted their confidence in {\name}'s descriptions. Other instances where {\name} failed to mention essential information also led to doubts:
\begin{quote}
\textit{``I am not confident because I put my eyedrop in front of me to see if it would pick it up, but it did not, which is fine as I guess it is not programmed for that. But it will be very useful in this case.''} - P1
\end{quote}
However, some pointed out the walk-up-and-use study design made them unable to fully explore and get used to {\name}: 
\begin{quote}
\textit{``Honestly, I remain conservative using it off the street tomorrow. But being used to the systems, I think if I had some time to get used to it, I could work with it.''} - P4
\end{quote}
But in general, they foresaw the promise and benefits {\name} can bring otherwise unavailable from existing apps, such as the real-time experiences and the adaptive level of details:
\begin{quote}
\textit{``It's closer to SeeingAI and BeMyEyes descriptive mode. Initially, it's like desk, chair, ... and become descriptive like a human, more color and context, if you are looking at things longer''} - P2
\end{quote}

\subsubsection{Perceived quality and customized visual information on the AI-generated descriptions}\label{quality}
\textbf{Participants found {\name} descriptions useful with adaptive and customized visual information, but felt overwhelmed in some situations. }
Participants noted several useful aspects of {\name}'s descriptions. For instance, {\name} starts with an overview and provides details on the fly for each new visual scene. 
Hence, if a participant's quick movements or turns lead to a succession of new scenes, they receive an overview for each. In contrast, if they focus on the same scene for a while, they receive detailed descriptions.
One participant noted:
\begin{quote}
    \textit{``It's interesting when it just provided only a few words when I moved or turned, like a desk, a chair, a person, it's nice to know what is included in this space. And I got details if I faced that for a little longer. I like the switch between these low-level and high-level descriptors. If I'm in the moment that I should picture things myself; it'll just give me low-level descriptors. I appreciate that ... But if I'm looking for something and I'm trying to figure out where I'm near, or get some landmarks and stuff. Then I appreciate the higher-level stuff.''} - P4
\end{quote}
Aside from the level of granularity, participants also perceived the increased descriptions in their customized visual attributes. For instance, P2 made our system verbose on color and spatial relationships and remarked:
\begin{quote}
    \textit{``This session did a better job at giving color descriptions. Also, it described more things like I said, location of things like in front of you, next to you, behind, you know, to your right or things of left.''} - P2
\end{quote}
Moreover, participants found {\name} offered unique and enriching experiences, \textit{``[{\name}] used strong words, so beautiful.''} (P3). They (P2, P4, P6) also pondered the balance between {\name}'s detailed visual descriptions and their practical use, suggesting that the descriptions should be more colloquial to mimic a human describer who provides the minimum viable information:
\begin{quote}
    \textit{I've never thought of a building being lit by tubes like a pattern or a line. It's all interesting for a blind person to have their eyes open to this stuff because I've never seen it before. It's all interesting information for me, but as far as practical use, I could get overwhelmed with it. Part of my brain loves it. Part of my brain is, Oh, I don't need it. So it's really interesting to be in this position. It really depends on the environment or your goal.} - P4
\end{quote}

\subsubsection{Alignment between users' mental model on the real world and what {\name} sees.}
\label{misalign}
\textbf{Participants desired the descriptions responsive to their physical reach, and the spatial information should center on them but not the image.}
During study sessions, we found that camera aiming issues caused misalignment between what users thought and what {\name} described. For instance, some participants (P2, P3) held the smartphone in their hands to explore their surroundings.
They were confused if {\name} missed describing the objects they touched, perceiving them to be something in their front captured by the camera. 
P2 frequently questioned during the study \textit{``It did not describe when I touched it [laptop]. I wasn't sure if I was getting it within the camera.'' } P4, who thought the camera did not capture what he needed, wanted to change the camera mount \textit{``Maybe next time I can use another camera headset that is over my eyes''}. 
Our subsequent video analysis found that their hands and the touched objects were beneath the smartphone and not captured within the frame.

Some participants also mentioned that {\name} could supplement their mobility with a white cane. For instance, they could confirm they had arrived at the exit upon hitting a chair, as {\name} had previously described the ``exit'' and ``chair'' together.:
\begin{quote}
    \textit{I have my cane and am able to follow the directions to the exit. When I go over that area, you see where the stack of chairs is, as [WorldScribe] mentioned that before. So like it would say chairs when I hit it. Okay I'm going the right way.
    } - P1
\end{quote}
However, what {\name} saw was based on a single image, limiting its understanding of the user's surroundings and creating erroneous spatial information.
For example, the `left' or `right' was determined by the spatial relationships within an image rather than the user's point of view. 
Some participants observed this discrepancy but understood that {\name} focused on describing their surroundings instead of providing directions. 
Consequently, they proposed integrating {\name} with other apps to provide more comprehensive experiences, which we described in the next section.

\subsubsection{Desires on more concrete information for practical use}\label{moreinfo}
\textbf{Participants desired concrete information for practical use in their daily lives, such as distance, directions, or pre-loaded map information.}
Participants found {\name} useful for general environmental understanding. However, for high-stakes scenarios such as navigation, participants believed {\name} could supplement their experiences with existing navigation apps. 
For instance, navigation apps often provide general directions such as \emph{``turn right at the next intersection''}, but it can be hard for BVI individuals to determine if they have reached the intersection or are at the crossroads. {\name} can assist this by describing their visual surroundings.
Additionally, some asked for more concrete information along with visual descriptions, such as spatial relationships from their perspective, exact distances to objects, and their continuous updates: 
\begin{quote}
\textit{``
    It would be like pre-loading the space. when I was looking for a classroom for approximately 3 doors or how many feet, and then you're gonna turn right. And then 4 doorways down on the right, that gives me a directional type of thing. When you're blind you have to do it by calling and checking the space, having that description and that context can help.''
} - P2
\end{quote}
Additionally, they wanted to have more control over the descriptions and integrate with spatial audio, as one participant, who used SoundScape \cite{soundscape}, mentioned:
\begin{quote}
    \textit{``I could have [{\name}] running in the background. It'd be almost like a lucid dream if you had it on the spatial audio. Okay, that's over there. I want to know more about that. So I turn toward it. Then, it changes the environment to show me that I'm facing that exact thing. That'd be really beautiful.''} - P4
\end{quote}
Although {\name} was not designed for real-world navigation, these insights from users are invaluable for guiding our next steps by designing and making visual descriptions integrated into more practical and high-stake scenarios.
We discuss the lessons learned from the study and potential improvements of live visual descriptions in Section \ref{discussion}.

\begin{figure*}[t]
\begin{center}
\includegraphics[width=\linewidth]{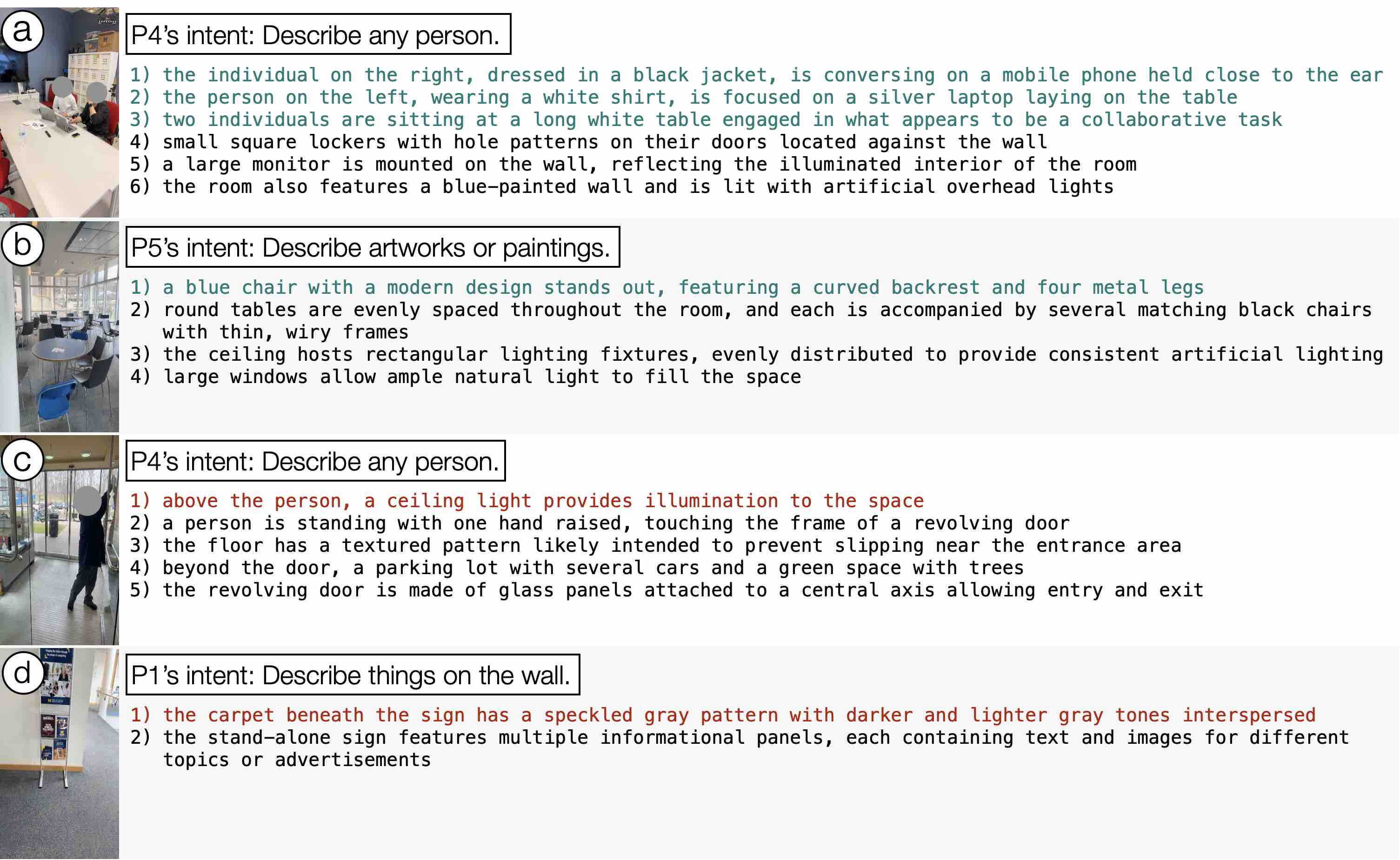}
\vspace{-1.8pc}
\caption{
Example results of description priority. 
(a) The descriptions relevant to the user's intent were successfully prioritized. 
(b) If no subjects meet the user's intent, descriptions would be ordered based on the distance to the user.
(c) The priority of similarity to the user's intent failed if a description involves the user's intent as supplement information such as spatial relationship.
(d) The distance priority to the user failed if the camera was angled down and {\name} took the floor as nearest to the user. 
}
\label{fig:priority_examples}
\Description{Figure 10
Example results of description priority. 
Figure 10 a 
The descriptions relevant to the user's intent were successfully prioritized. Image shows texts, including P4’s intent: Describe any person. (1) the individual on the right, dressed in a black jacket, is conversing on a mobile phone held close to the ear. (2) the person on the left, wearing a white shirt, is focused on a silver laptop laying on the table. (3) two individuals are sitting at a long white table engaged in what appears to be a collaborative task. (4) small square lockers with hole patterns on their doors located against the wall. (5) a large monitor is mounted on the wall, reflecting the illuminated interior of the room. (6) the room also features a blue-painted wall and is lit with artificial overhead lights
Figure 10 b 
If no subjects meet the user's intent, descriptions would be ordered based on the distance to the user. Image shows texts, including P5’s intent: Describe artworks or paintings. (1) a blue chair with a modern design stands out, featuring a curved backrest and four metal legs. (2) round tables are evenly spaced throughout the room, and each is accompanied by several matching black chairs with thin, wiry frames.
(3) the ceiling hosts rectangular lighting fixtures, evenly distributed to provide consistent artificial lighting. (4) large windows allow ample natural light to fill the space.
Figure 10 c 
The priority of similarity to the user's intent failed if a description involves the user's intent as supplement information such as spatial relationship. The image shows texts, including P4’s intent: Describe any person. (1) above the person, a ceiling light provides illumination to the space. (2) a person is standing with one hand raised, touching the frame of a revolving door. (3) the floor has a textured pattern likely intended to prevent slipping near the entrance area. (4) beyond the door, a parking lot with several cars and a green space with trees. (5) the revolving door is made of glass panels attached to a central axis allowing entry and exit
Figure 10 d 
The distance priority to the user failed if the camera was angled down and WorldScribe took the floor as nearest to the user. The image shows texts, including P1’s intent: Describe things on the wall. (1) the carpet beneath the sign has a speckled gray pattern with darker and lighter gray tones interspersed. (2) the stand-alone sign features multiple informational panels, each containing text and images for different topics or advertisements.}
\end{center}
\end{figure*}

\section{Pipeline Evaluation}\label{technical_evaluation}
In this evaluation, we measured the accuracy, coverage of user-desired content, and description priority based on users' intent and proximity of described content.
We collected data from our user evaluation, such as descriptions and their generative models, timestamps, referenced frames and prompts, customization settings, screen recordings, etc.
These videos and frames were naturally captured by BVI participants in our user evaluation, resulting in camera motions or slants that impacted image quality, but it preserved the authenticity and relevance of our findings to real-world experiences.
Each study session's video recording lasted around ten minutes, as described in Section \ref{user_evaluation}.

\subsection{Accuracy}
We measured the accuracy of {\name} descriptions by inspecting the description content and the referenced frame.

\textbf{Dataset \& Analysis.}
In total, we collected 2,350 descriptions from our user study.
The description sources included YOLO World \cite{yolow}, Moondream \cite{moondream}, and GPT-4v \cite{gpt4v}.
We inspected each description and considered descriptions incorrect if they could not be justified through their referenced camera frame.

\textbf{Results.}
Overall, we found that 370 of 2,350 instances (15.74\%) were incorrect in relation to the referenced camera frame. 
Specifically, 122 of the 638 descriptions (19.12\%) from YOLO World were incorrect, 72 of the 549 (13.11\%) descriptions from Moondream were incorrect, and 176 of the 1,157 (15.21\%) descriptions from GPT-4v \cite{gpt4v} were incorrect.

We observed several reasons for incorrect descriptions. 
For instance, when prompted to generate object classes based on users' intent for YOLO World \cite{yolow}, GPT-4 \cite{gpt4} sometimes generated classes that could not be identified in our study environment, such as \textit{museum}, \textit{exhibition}, and \textit{classroom}. Additionally, the low image quality significantly impacted the accuracy of descriptions. For example, YOLO World \cite{yolow} often mistook whiteboards, paper, walls, or illuminated monitor screens for other objects.
Moondream \cite{moondream}, prompted to provide general descriptions, performed well in covering common objects and their spatial relationships but sometimes included hallucinated content.
For GPT-4v \cite{gpt4v}, lighting conditions or capture angles affected the results. For example, a cabinet was mistaken for a washer and dryer when shot from the side, cluttered folding chairs were mistaken for motorcycles or bicycles, and a male with long hair was identified as female.

\subsection{Coverage of User Intent}\label{coverage}
We measured if {\name} descriptions covered important content of the users' intent in a timely fashion.

\textbf{Dataset \& Analysis. }
We used the smartphone video recordings of the P1-P5 self-defined scenario due to their concrete intent.
To determine whether the descriptions covered the essential content related to users' specified intent, we developed video codes to annotate the objects that should be described in the footage. 
One of the authors carefully reviewed each recording and annotated the objects relevant to users' intent. For example, we labeled items on the wall for P1's intent \PROMPT{Describe things on the wall} and labeled people for P4's intent \PROMPT{Describe any person.} 
In each video, we observed that participants sometimes turned around or moved frequently, or people in the video also moved dynamically, leaving some objects of interest to appear only briefly in the video. 
Thus, we annotated an object only if it lasted long enough and was clear enough to identify without pausing the video. 
Each label covered a time range from when the object appeared to when it disappeared from view. 
Another author then examined each label and marked it as \textit{covered} if the descriptions within the time range included the annotated objects. 
We had 64 labels in total from the five video recordings.

\textbf{Limitations.}
Due to the small pool of participants and the limited study time, we did not have a comparable number of ground truth labels to those standard video datasets. We will discuss more about this in Section \ref{benchmark}.

\textbf{Results.}
We found that 75\% of annotated objects (48 out of 64) were covered by {\name} descriptions. 
Based on our post examination, {\name} successfully described and covered objects of users' intent in most cases if users faced for a long period, but failed in a few cases that an object was partially occluded.
We also observed that {\name} failed to describe an object of users' intent in time if {\name} was still describing the previous visual scene while users had already moved to a new one.

\subsection{Description Priority}
We measured if {\name} prioritized descriptions based on users' intent or the proximity of described content to users.

\textbf{Dataset \& Analysis. } 
In total, we collected 120 descriptions by randomly selecting 20 descriptions generated by GPT-4v \cite{gpt4v} from each participant's self-defined scenario.
We marked an instance as correct if the presented description was relevant to the user's intent, or if the described content was nearest to the user.

\textbf{Results.}
Our analysis revealed that 97 out of 120 descriptions (80.83\%) aligned with user intent or were prioritized based on the proximity of the described content to the user. 
We found that errors commonly occurred when the relevant information was present but not the focus, such as considering the user's intended object as spatial reference (Figure \ref{fig:priority_examples}c). 
Additionally, camera angles often varied, sometimes tilting towards the floor or ceiling, which was recognized as the nearest content to the user (Figure \ref{fig:priority_examples}d).

\section{Discussion and Future Work}\label{discussion}
In this section, we discuss our lessons learned and design implications for context-aware and customizable live visual descriptions.

\subsection{Challenges in Describing the Real World}\label{morechallenges}
Describing the real world is more challenging than digital visual media due to the need for timely descriptions aligned with users' intent and the higher standards and expectations in high-stakes situations. 
While {\name} made an important step toward providing context-aware live descriptions, participants brought up several aspects and challenges for future research to address.

First, while {\name} simulated short-term memory by avoiding repeated descriptions within preceding sentences (Section \ref{select_new_description}), participants expressed a need for more sophisticated long-term memory for visual descriptions. They suggested that previously navigated spaces or paths should not be reintroduced upon revisiting; instead, only visual changes or new elements since the last visit should be described. Spatial information should reference a series of camera frames or more complete data source to construct and represent users' environment (\textit{e.g.,} real-time NeRF \cite{li2022rtnerf,duckworth2023smerf}), rather than relying on a single video frame.

Second, it is hard for users to express their intent in a few sentences, which may implicitly change over time when exploring an area. 
\CHANGE{This need to update intents was highlighted by our participants, who hoped to converse with the system to update their intent or clarify the confusing descriptions, similar to how they interact with human describers. 
Ideally, aside from such turn-by-turn interactions,} a context-aware live visual description system should implicitly learn and adapt to the user's intent and environment through long-term interactions with users to reduce friction and increase usability.
Future works could incorporate other data sources and modalities, such as GPS data, maps, visual details in videos or images, and description history to enable long-term memory.

\subsection{Towards More Humanized Descriptions}
Besides the challenges in crafting useful descriptions for the real world, the way to present descriptions could also influence users' understanding or engagement with visual media or scenes. For instance, describing from a first-person or third-person perspective could affect immersion in the environment \cite{omniscribe}. Tone, voice, and syntax \cite{ndr,adp,acb_guideline,3PlayMedia_BeginnerGuideline, DCMP, mac} could also significantly impact experiences and comprehension.
During our study, we received varied comments and preferences on these presentation aspects. For example, participant P4 described {\name}'s voice as "hoarse," making the content unclear and uncomfortable. 
While some appreciated {\name}'s current tone, others found the descriptions "artistic" or "poetic" and too wordy for practical use.
Also, while participants appreciated {\name}'s pauses or increased volume for presentation clarity, they hoped {\name} could provide transitioning descriptions or earcons when shifting to a new visual scene to ensure it was describing the current but not the previous scene. 
They further noted that human describers use more colloquial language than {\name} when conveying useful information, and prioritize their clarity over grammatical nuances. 
Future works should consider and enable more customizations of presentation.

\subsection{Benchmarking Dataset for Live Descriptions}\label{benchmark}
Our pipeline evaluation was limited to data from six BVI people to reflect real-world experiences with {\name}, which are different from the current video captioning dataset in several aspects.

First, the quality of video is quite different from that of standard datasets.
For instance, frames occasionally appear at unusual angles or become blurred due to several factors such as users' movement, whether the camera is handheld, attached to a swinging lanyard, or tucked inside a pocket.
Second, objects of interest may not appear consistently across consecutive frames and could be partially obscured or located to the periphery, as camera aiming is particularly challenging for BVI people \cite{helpaiming2012,visphoto,vizwiz,socialvizwiz,help2014aiming}.
Third, to provide useful live visual descriptions, it is important to provide concrete details beyond describing visual events, such as providing distances or sizes in concrete units (\textit{e.g.,} feet, meters).
Fourth, spatial relationships of objects should pivot to users' perspective (\textit{e.g.,} using clock directions), rather than the image itself (\textit{e.g.,} something on your left but not something on the left of the image).

To enable appropriate evaluation of live descriptions, additional datasets and metrics are needed.
First, a potential metric is \textit{contextual responsiveness}, which evaluates if the utility of descriptions aligns with the current context, such as having directions during navigation, having rich adjectives when viewing artworks, or describing objects reached by the user physically. 
The second potential metric is \textit{contextual timeliness}. For instance, high-stakes scenarios may require a higher timeliness to signal potential danger before it happens (\textit{e.g.,} the status of traffic light, whizzing car), while low-stakes scenarios could have much room for latency. 
Third, a potential metric is \textit{contexual detailedness}, which evaluates whether a description provides only the necessary information without excess visual detail (\textit{e.g.,} using multiple adjectives when the user is only interested in color, or describing the status of all three traffic lights instead of just the lit one).
Overall, evaluating the context-awareness of live descriptions involves multiple factors. 
To fulfill such high demand for live visual descriptions in the real world, future works should develop ways to collect and annotate video datasets shot by BVI people. 
\CHANGE{It is also notable for including such datasets in existing ones to carefully build a universal and unbiased dataset that is not skewed toward any particular group. }

\subsection{Generalizing {\name}}
We envision expanding {\name} to other media formats and integrating the rapidly evolving AI capabilities in the future. 
First, {\name} could be tailored to visual media that require immediate descriptions.
For instance, when describing 360-degree videos \cite{omniscribe}, it was hard for describers to pre-populate audio descriptions for the different fields of view with rich information due to the user's unpredictable viewing trajectory. 
It would be beneficial for {\name} to generate live descriptions responsive to the user's current view, while automatically pausing descriptions when important sounds happen (\textit{e.g.,} narration), or increasing volume when unimportant sounds occur (\textit{e.g.,} background music) in the video. 

\CHANGE{Second, {\name} could also be extended to support low-vision users who may use wearables to receive visual aids in the real world \cite{cuesee,foresee,flexisee}. 
For instance, the type of visual enhancement could be determined based on the user's mobility states and visual scenes, similar to {\name} detecting the user's orientation and frame similarity to provide corresponding descriptions. 
Also, live visual descriptions could confirm the visual scene for low-vision users, and {\name} could use the visually enhanced image frame from the wearables to increase description accuracy. 
Future works could explore integrating different assistive technologies (\textit{e.g.,} wearables, navigation systems) as an ecosystem to provide corresponding contextual support.
}

\subsection{Directions with Rapid Evolution of Future Large Models}
{\name} provided live visual descriptions by leveraging LLMs to understand users' intent and an architecture that balances the tradeoffs between latency and richness of different VLMs. 
Given the rapid evolution of VLMs and LLMs and computing power in recent years, it is foreseeable that accuracy and latency will significantly improve.
\CHANGE{This progress may possibly lead to the reliance on fewer or even a single large multimodal model (\textit{e.g.,} GPT-4o \cite{gpt4o}) to generate descriptions of varying granularity, but raising further questions about which inputs beyond images should be included when prompting. 
Additional contextual factors, such as environmental sounds, GPS data, and users' state and activities, could be considered, and the prompt structure could be dynamically changed based on these inputs and user needs.}
Future work could also explore incorporating more advanced AI models into the description generation process. 
For instance, we could improve descriptions with additional verification \cite{genassist}, deblur or increase resolution of the image for clarity, construct a 3D scene on the fly to provide accurate spatial information \cite{li2022rtnerf,duckworth2023smerf}, and explain sound causality by cross-grounding visual and audio data \cite{tian2021cyclic, zhao2023bubogpt, crossa11y}. 
Future works should explore these possibilities in such a rapidly evolving landscape of computational platforms and AI model capabilities.

\section{Conclusion}
We have presented {\name}, a system towards providing context-aware live visual descriptions to facilitate the environmental understanding for BVI people. 
Through a formative study with five BVI people, we identified several design goals of providing context-aware live visual descriptions. 
We implemented several components to tailor user's contexts, 
such as enabling users to specify their intent and generate descriptions tailored to their needs, providing consecutive short or long detailed descriptions based on visual context, and presenting descriptions with pausing or volume increased based on the sound context.
Through an evaluation with six BVI people, we demonstrated how they perceived the {\name} descriptions and identified gaps in fulfilling their expectations for using {\name} descriptions in practice.
Through a pipeline evaluation, we showed {\name} can provide fairly accurate visual descriptions, cover information about the user's intent, and prioritize descriptions based on the user's intent.
Finally, we discussed more challenges in describing the real world, how to make descriptions more humanized and usable, and potential benchmark datasets. 
Through this work, we also recognized promoting real-world accessibility through live descriptions will be a long-term overarching problem, considering the diversity of people's needs and the complexity of real-world environments.

\begin{acks}
We thank our anonymous reviewers and all the participants in our study for their suggestions, as well as Andi Xu for helping facilitate our user studies.
\end{acks}

\bibliographystyle{ACM-Reference-Format}
\bibliography{worldscribe}

\newpage
\clearpage
\appendix

\begin{figure*}[!t]
\vspace{0pc}
\begin{center}
\includegraphics[width=\linewidth]{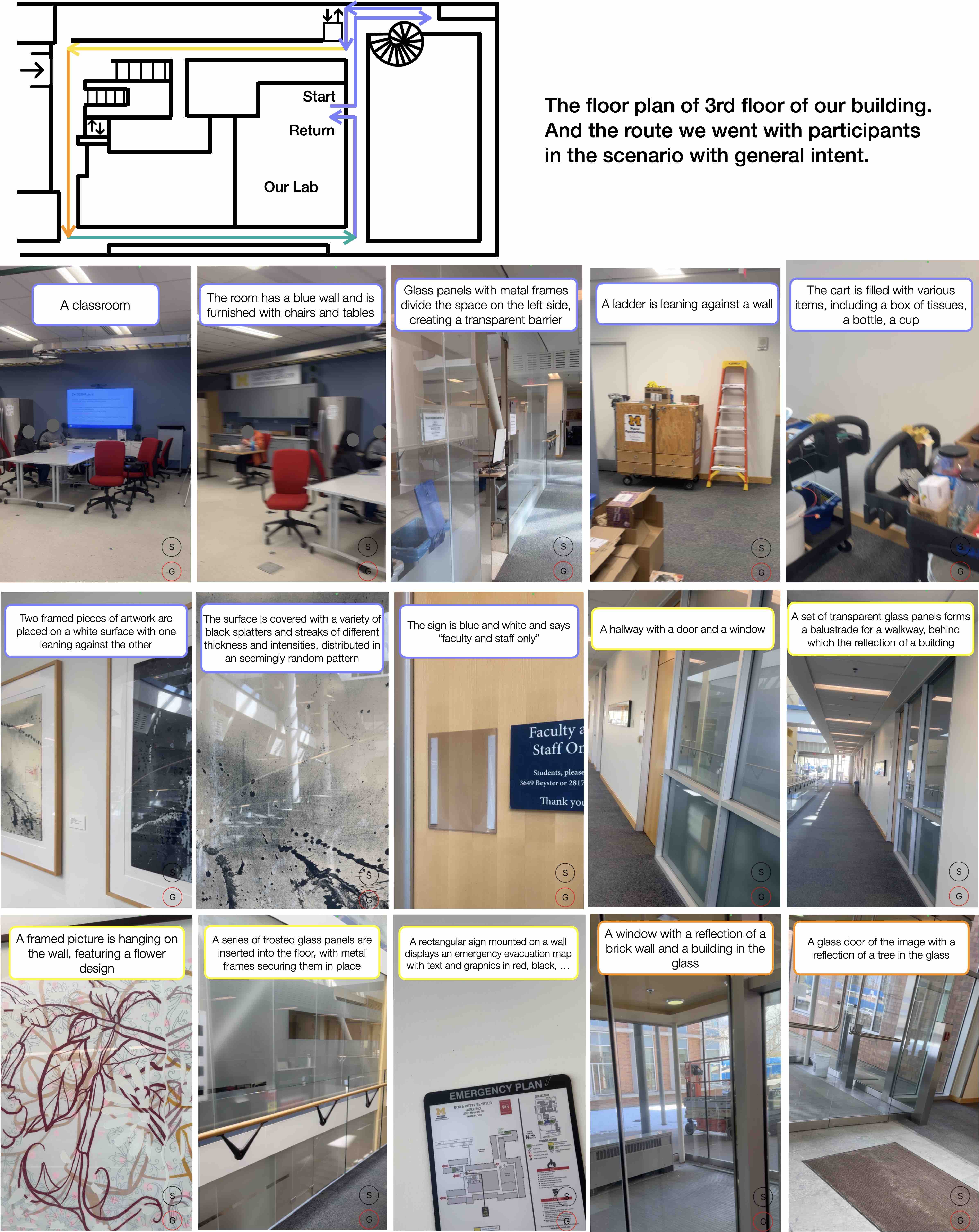}
\vspace{-1.pc}
\label{fig:test}
\Description{Example setup and descriptions in the study with P6}
\end{center}
\end{figure*}

\begin{figure*}[t]
\vspace{-0.5pc}
\begin{center}
\includegraphics[width=\linewidth]{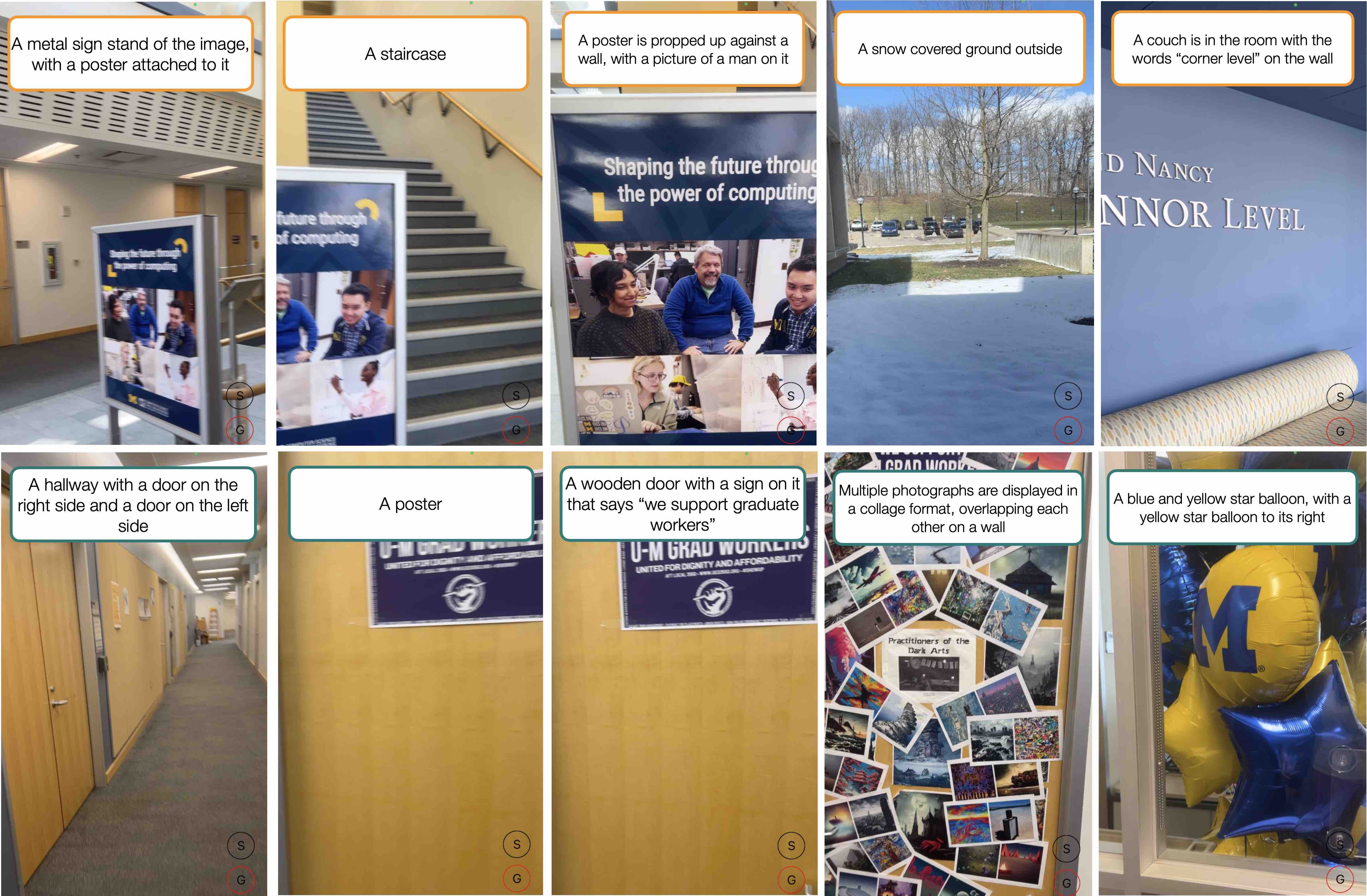}
\caption{Example study route, keyframes, and WorldScribe-generated descriptions in our user study with P6.}
\label{fig:test2}
\Description{Figure 11
The floor plan of 3rd floor of our building. And the route we went with participants in the scenario with general intent. Descriptions from top left to bottom right of the figure are:
(1) A classroom.
(2) The room has a blue wall and is furnished with chairs and tables.
(3) Glass panels with metal frames divide the space on the left side, creating a transparent barrier.
(4) A ladder is leaning against a wall.
(5) The cart is filled with various items, including a box of tissues, a bottle, a cup.
(6) Two framed pieces of artwork are placed on a white surface with one leaning against the other.
(7) The surface is covered with a variety of black splatters and streaks of different thickness and intensities, distributed in an seemingly random pattern.
(8) The sign is blue and white and says “faculty and staff only”.
(9) A hallway with a door and a window.
(10) A set of transparent glass panels forms a balustrade for a walkway, behind which the reflection of a building.
(11) A framed picture is hanging on the wall, featuring a flower design.
(12) A series of frosted glass panels are inserted into the floor, with metal frames securing them in place.
(13) A rectangular sign mounted on a wall displays an emergency evacuation map with text and graphics in red, black, …
(14) A window with a reflection of a brick wall and a building in the glass.
(15) A glass door of the image with a reflection of a tree in the glass.
(16) A metal sign stand of the image, with a poster attached to it.
(17) A staircase. 
(18) A poster is propped up against a wall, with a picture of a man on it. 
(19) A snow covered ground outside.
(20) A couch is in the room with the words “corner level” on the wall.
(21) A hallway with a door on the right side and a door on the left side.
(22) A poster.
(23) A wooden door with a sign on it that says “we support graduate workers”.
(24) Multiple photographs are displayed in a collage format, overlapping each other on a wall.
(25) A blue and yellow star balloon, with a yellow star balloon to its right.
}
\end{center}
\end{figure*}

\end{document}